 \documentclass[final,3p,times]{elsarticle}

\usepackage{graphicx, amssymb, epsfig, adjustbox, amsthm, xcolor, amsmath,natbib, booktabs, bigints,soul}
\usepackage{hyperref}

\setstcolor{red}

\numberwithin{equation}{section}
\theoremstyle{plain}
\newtheorem{thm}{Theorem}[section]

\hypersetup{pdfauthor={Name}}

\begin{document}

\begin{frontmatter}


\title{Normalized Power Prior Bayesian Analysis}



\author[utsa]{Keying Ye}
\author{Zifei Han\corref{}\fnref{uibe}}
\ead{zifeihan@uibe.edu.cn; zifei_han@outlook.com;keying.ye@utsa.edu}
\author[novartis]{Yuyan Duan}
\author[fda]{Tianyu Bai}
\address[utsa]{Department of Management Science and Statistics, 
The University of Texas at San Antonio, San Antonio, TX, USA.}

\address[uibe]{School of Statistics, 
University of International Business and Economics, Beijing, China.}
\address[novartis]{Novartis Institutes for BioMedical Research, Cambridge, MA, USA.}
\address[fda]{U.S. Food and Drug Administration, Silver Spring, MD, USA}

\begin{abstract}
The elicitation of power priors, based on the availability of historical data, 
is realized by raising the likelihood function of 
the historical data to a 
fractional power $\delta$, 
which quantifies the degree of discounting of the 
historical information in making 
inference with the current data. 
When $\delta$ is not pre-specified and is treated as random, 
it can be estimated from the data 
using Bayesian updating paradigm. 
However, in the  original form of the joint power prior Bayesian approach,
certain positive constants before the likelihood of the 
historical data could be multiplied when different settings of sufficient statistics are employed. 
This would change the power priors with different constants, 
and hence the likelihood principle is violated.

In this article, we investigate 
a normalized power prior approach which obeys the likelihood principle and is a modified form of the joint power prior. 
The optimality properties of the 
normalized power prior in the sense of 
minimizing the weighted Kullback-Leibler divergence is investigated.
By examining the posteriors of several commonly used  distributions, 
we show that the discrepancy between the historical and the current data can be well quantified by the power parameter under the normalized power prior setting. 
Efficient algorithms to compute the scale factor is also proposed. 
In addition, we illustrate the use of the normalized power prior Bayesian analysis with three data examples, and 
provide an implementation with an R package NPP.
\end{abstract}

\end{frontmatter}

\section{Introduction}

In applying statistics to real experiments, it is common that the
sample size in the current study is inadequate to provide
enough precision for parameter estimation, while plenty of the 
historical data or data from similar research settings 
are available. 
For example, when design a clinical study, 
historical data of the standard care might be available from 
other clinical studies or a patient registry. 
Due to the nature of sequential information updating, it is natural
to use a Bayesian approach with an informative prior on the model
parameters to incorporate these historical data. 
Though the current and historical data
are usually assumed to follow distributions from the same family, the
population parameters may change somewhat over different time and/or experimental settings. 
How to adaptively incorporate the historical data considering the data
heterogeneity becomes a major concern for the 
informative prior elicitation. 

To address this issue, \cite{IbrahimChen98}, and
thereafter \cite{Chen00}, \cite{IbrahimChen00}, and 
\cite{Ibrahim03} proposed the concept of {\em power priors},  
based on the availability of historical data. 
The basic idea is to raise the likelihood function based on 
the historical data to a {\em power parameter} $\delta$  
$(0\leq\delta\leq1)$ that controls the influence of the 
historical data. 
Its relationship with hierarchical models is also shown by \cite{Chen06}. 
For a comprehensive review of the 
power prior, we refer the readers to the 
seminar article \cite{Ibrahim15}. 
The power parameter $\delta$ can be prefixed according to   
external information. 
It is also possible to search for a reasonable level of information 
borrowing from the prior-data conflict via  
sensitivity analysis 
according to certain criteria. 
For example, \cite{Ibrahim12} suggested the use of 
deviance information criterion \citep{Spiegelhalter02} or the 
logarithm of pseudo-marginal likelihood. 
The choice of $\delta$ would depend on the criterion of interest.

\cite{IbrahimChen00} and \cite{Chen00} 
generalized the power prior with a fixed $\delta$ to 
a random $\delta$ by introducing the {\em joint power priors}. 
They specified a joint prior distribution 
directly for both $\delta$ and $\boldsymbol{\theta}$, the parameters in consideration, in which an independent proper 
prior for $\delta$ was considered in addition to the original form of the power prior.
Hypothetically, when the initial prior for $\delta$ is 
vague, the magnitude of borrowing would be mostly 
determined by the heterogeneity between the historical and the 
current data. 
However, under the joint power priors, the posterior distributions 
vary with the constants before the historical likelihood functions, 
which violates the likelihood principle \citep{Birnbaum62}. 
It raises a critical question regarding which likelihood 
function should be used in practice. 
For example, the likelihood function  
based on the raw data and the likelihood function based on the 
sufficient statistics could differ by a multiplicative constant. 
This would likely yield different posteriors. 
Therefore, it may not be appropriate \citep{Neuenschwander09}. 
Furthermore, 
the power parameter has a tendency to be close to zero empirically, 
which suggests that much of a 
historical data may not be used in decision making \citep{Neelon10}. 

In this article, we investigate a modified power prior which was 
initially proposed by \cite{Duan06a} for a random $\delta$. 
It is named as the {\it normalized power prior} since 
it includes a scale factor. 
The normalized power prior obeys the likelihood principle.  
As a result, the posteriors can quantify the compatibility between 
the current and historical data automatically, 
and hence control the influence of historical data on 
the current study in a more sensible way.

The goals of this work are threefold. 
First, we review the joint power prior and the  
normalized power prior that have been proposed in literature.  
We aim to show that the joint power prior 
may not be appropriate for a random $\delta$. 
Second, we carry out a comprehensive study on properties of the 
normalized power prior both theoretically and numerically, 
shed light on the posterior behavior in response to the  
data compatibility. 
Finally, we design efficient computational algorithms and  
provide practical implementations along with three data examples.

\section{A Normalized Power Prior Approach}\label{sec:theory}
\subsection{The Normalized Power Prior}\label{subsec:NewPrior}

Suppose that $\boldsymbol{\theta}$ is the parameter (vector or scalar) of interest and
$L(\boldsymbol{\theta}|D_{0})$ is the likelihood function of 
$\boldsymbol{\theta}$ based on 
the historical data $D_{0}$.  
In this article, we assume
that the historical data $D_{0}$ and current data $D$  
are independent random samples.  
Furthermore, denote by $\pi_{0}(\boldsymbol{\theta})$ the 
{\em initial prior} for $\boldsymbol{\theta}$. 
Given the power parameter $\delta$, \cite{IbrahimChen00} defined the 
{\em power prior} of $\boldsymbol{\theta}$ for the current study as
\begin{equation}
\pi(\boldsymbol{\theta}|D_{0},\delta) \propto
L(\boldsymbol{\theta}|D_{0})^{\delta}\pi_{0}(\boldsymbol{\theta}). \label{eq:originalprior}
\end{equation}
The power parameter $\delta$, a scalar in $[0, 1]$, measures the influence of historical information  on the current study. 

The power prior $\pi(\boldsymbol{\theta}|D_{0},\delta)$ in (\ref{eq:originalprior})
was initially elicited for a fixed $\delta$.  
As the value of $\delta$ is not necessarily pre-determined 
and typically unknown in practice, 
the full Bayesian approach extends the case 
to a random $\delta$ by assigning
a reasonable initial prior $\pi_{0}(\delta)$ on it. 
A natural prior for $\delta$ would be a 
$\text{Beta}(\alpha_\delta, \beta_\delta)$
distribution since $0\leq \delta \leq 1$.  
\cite{IbrahimChen00} 
constructed the {\it joint power prior} of 
$(\boldsymbol{\theta},\delta)$ as
\begin{equation}
\label{eq:jointpp}
    \pi(\boldsymbol{\theta},\delta|D_{0}) \propto 
    L(\boldsymbol{\boldsymbol{\theta}}|D_{0})^\delta \pi_{0}(\boldsymbol{\theta}) \pi_{0}(\delta), 
\end{equation}
with the posterior, given the current data $D$, as 
\begin{eqnarray}
\label{eq:originalpost} \pi(\boldsymbol{\theta},\delta|D_0,D)=
\frac{L(\boldsymbol{\theta}|D)
L(\boldsymbol{\theta}|D_0)^{\delta}\pi_{0}(\boldsymbol{\theta})
\pi_{0}(\delta)}{\int_{0}^{1}  \pi_{0}(\delta) \left\{
\int_{\boldsymbol{\Theta}}L(\boldsymbol{\theta}|D) L(\boldsymbol{\theta}|D_0)^{\delta}
\pi_{0}(\boldsymbol{\theta})
d\boldsymbol{\theta}\right\} d\delta}, 
\end{eqnarray}
where $\boldsymbol{\Theta}$ denotes the parameter space of $\boldsymbol{\theta}$. 
The prior in (\ref{eq:jointpp}) is constructed 
by directly assigning a prior for 
$(\boldsymbol{\theta}, \delta)$ jointly \citep{Ibrahim15}.
However, if we integrate $\boldsymbol{\theta}$
out in (\ref{eq:jointpp}) we have $
\pi(\delta|D_0) 
\propto \pi_{0}(\delta) \int_{\boldsymbol{\Theta}}
L(\boldsymbol{\theta}|D_{0})^\delta
\pi_{0}(\boldsymbol{\theta}) d\boldsymbol{\theta}$, 
which does not equal to $\pi_{0}(\delta)$. This meant that 
the initial prior for $\delta$ is updated 
after one observes the historical data alone. 
Moreover, in the posterior (\ref{eq:originalpost}), 
any constant before $L(\boldsymbol{\theta}|D_0)$
cannot be canceled out between the numerator and 
the denominator. 
This could yield different posteriors if different forms of the
likelihood functions are used. 
For example, the likelihood based on the raw data and 
the likelihood based on the distribution of 
sufficient statistics could 
result in different posteriors. 
Also, the prior in (\ref{eq:jointpp}) could be improper. 
Once the historical information is available, 
a prior elicited from such information would better be proper.
Propriety conditions for four commonly used classes of regression 
models can be found in \cite{IbrahimChen00} and \cite{Chen00}.

Alternatively, one can first specify a conditional prior distribution on  
$\boldsymbol{\theta}$ given $\delta$, then specify a marginal distribution for $\delta$. 
The normalizing constant in the first step is therefore a function of $\delta$. 
Since $\delta$ is a parameter, this scale factor 
$C(\delta) = \int_{\boldsymbol{\Theta}} L(\boldsymbol{\theta}|D_{0})^\delta \pi_{0}(\boldsymbol{\theta}) 
d \boldsymbol{\theta}$ should not be ignored. 
Therefore, a modified power prior formulation, called the 
{\it normalized power prior}, was proposed by 
\cite{Duan06a} which included this scale factor. 
Consequently, for $(\boldsymbol{\theta},\delta)$, the normalized power prior is
\begin{equation}
\pi(\boldsymbol{\theta},\delta|D_{0}) \propto 
\frac{L(\boldsymbol{\theta}|D_{0})^\delta
\pi_{0}(\boldsymbol{\theta})\pi_{0}(\delta)}{\int_{\boldsymbol{\Theta}}
L(\boldsymbol{\theta}|D_{0})^\delta
\pi_{0}(\boldsymbol{\theta}) d\boldsymbol{\theta}}, 
\label{eq:powerprior}
\end{equation}
in the region of $\delta$ such that the denominator of
(\ref{eq:powerprior}) is finite.

When $\int_{\boldsymbol{\Theta}} L(\boldsymbol{\theta}|D_{0})^{\delta}
\pi_{0}(\boldsymbol{\theta})  d \boldsymbol{\theta} <\infty$, 
the prior in (\ref{eq:powerprior}) is always proper 
given that $\pi_{0}(\delta)$ is proper, 
whereas it is not necessarily
the case for that of the joint power prior (\ref{eq:jointpp}). 
More importantly,
multiplying the likelihood function in (\ref{eq:jointpp}) by an
arbitrary positive constant, which could be a function of $D_0$, may change the joint power prior, whereas the constant
is canceled out in the normalized power prior in  (\ref{eq:powerprior}). 

Using the current data to update the prior distribution
$\pi(\boldsymbol{\theta},\delta|D_{0})$ in (\ref{eq:powerprior}),
we derive the joint posterior distribution for $(\boldsymbol{\theta}, \delta)$ as
\begin{equation*}
\pi(\boldsymbol{\theta},\delta|D_{0},D) \propto
L(\boldsymbol{\theta}|D)\pi(\boldsymbol{\theta},\delta|D_{0}) \propto
\frac{L(\boldsymbol{\theta}|D)L(\boldsymbol{\theta}|D_{0})^\delta 
\pi_{0}(\boldsymbol{\theta})
\pi_{0}(\delta)}{\int_{\boldsymbol{\Theta}} L(\boldsymbol{\theta}|D_{0})^\delta 
\pi_{0}(\boldsymbol{\theta})d\boldsymbol{\theta}}.
\end{equation*}
Integrating $\boldsymbol{\theta}$ out from the expression above, the marginal
posterior distribution of $\delta$ can be expressed as
\begin{equation}
\pi(\delta|D_{0},D) \propto 
\pi_{0}(\delta) \frac{\int_{\boldsymbol{\Theta}}
L(\boldsymbol{\theta}|D)L(\boldsymbol{\theta}|D_{0})^\delta 
\pi_{0}(\boldsymbol{\theta})  d\boldsymbol{\theta}}
{\int_{\boldsymbol{\Theta}} L(\boldsymbol{\theta}|D_{0})^\delta \pi_{0}(\boldsymbol{\theta}) 
d\boldsymbol{\theta}}.
\label{aposterior} 
\end{equation}
If we integrate $\delta$ out in (\ref{eq:powerprior}), we obtain a
new prior for $\boldsymbol{\theta}$, a prior that is updated by the historical
information,
\begin{equation}\label{pprior}
\pi(\boldsymbol{\theta}|D_0) = \int_{0}^{1} \pi
(\boldsymbol{\theta},\delta|D_0) d\delta \propto
\pi_{0}(\boldsymbol{\theta})
\int_{0}^{1} \frac{L(\boldsymbol{\theta}|D_{0})^\delta
\pi_{0}(\delta)
}{\int_{\boldsymbol{\Theta}} L(\boldsymbol{\theta}|D_{0})^\delta \pi_{0}(\boldsymbol{\theta}) d\boldsymbol{\theta}} 
d\delta.
\end{equation}

With historical data appropriately incorporated, (\ref{pprior})
can be viewed as an informative prior for the Bayesian analysis to
the current data.  Consequently, the posterior distribution of
$\boldsymbol{\theta}$ can be written as
\begin{equation*} 
\pi(\boldsymbol{\theta}|D_0,D) \propto \pi(\boldsymbol{\theta}|D_0)
L(\boldsymbol{\theta}|D) \propto
\pi_{0}(\boldsymbol{\theta})L(\boldsymbol{\theta}|D)
\int_{0}^{1} 
\frac{L(\boldsymbol{\theta}|D_{0})^\delta
\pi_{0}(\delta)
}{\int_{\boldsymbol{\Theta}} L(\boldsymbol{\theta}|D_{0})^\delta \pi_{0}(\boldsymbol{\theta}) 
 d\boldsymbol{\theta}} d\delta. 
\end{equation*}

Below we describe some variations of the normalized power prior. 
A primary extension deals with the presence of multiple historical studies.
Similar to \cite{IbrahimChen00}, 
the prior defined in (\ref{eq:powerprior}) can be 
easily generalized. 
Suppose there are $m$ historical studies,  
denote by $D_{0j}$ the historical data for the $j^{th}$ study,
$j=1,\ldots,m$ and 
$\boldsymbol{D}_{0}=(D_{01},\ldots,D_{0m})$. 
The power parameter for each historical study can be different, 
and we can further assume they follow the same independent initial 
prior. 
Let $\boldsymbol{\delta}=(\delta_{1},\ldots,\delta_{m})$, 
the normalized power prior of the form (\ref{eq:powerprior}) 
can be generalized to  
\begin{equation*}
\pi(\boldsymbol{\theta},\boldsymbol{\delta}|
\boldsymbol{D}_{0}) \propto
\frac{\bigg\{\prod_{j=1}^{m}L(\boldsymbol{\theta}|D_{0j})^{\delta_{j}}
\pi_{0}(\delta_{j})\bigg\} \pi_{0}(\boldsymbol{\theta})}{\displaystyle
\int_{\boldsymbol{\Theta}}
\bigg\{\prod_{j=1}^{m}L(\boldsymbol{\theta}|D_{0j})^{\delta_{j}}\bigg\}
\pi_{0}(\boldsymbol{\theta})
 d\boldsymbol{\theta}}.
\end{equation*}

This framework would accommodate the 
potential heterogeneity among
historical data sets from different sources or collected at different time 
points.
Data collected over a long period may be divided into
several historical data sets to ensure the homogeneity within each
data. 
Examples of implementing the 
power prior approach using multiple historical studies can
be found in \cite{Duan06a}, \cite{Gamalo14}, \cite{Gravestock18} and \cite{Banbeta19}. 

An important extension is based on the 
{\it partial borrowing power prior} \citep{Ibrahim12-2, Chen14-2}, 
in which the historical data can be borrowed 
only through some common parameters with fixed 
$\delta$. 
For instance, when evaluating cardiovascular risk in new therapies, 
priors for only a subset of the parameters are 
constructed based on the historical data 
\citep{Chen14}. 
Below we describe the {\it partial borrowing normalized power prior},
which is an extension of the partial borrowing power prior. Let   
$\boldsymbol{\theta} = (\boldsymbol{\theta}_{c}, \boldsymbol{\theta}_{1})$ 
be the parameter of interest in the current study, and let 
$(\boldsymbol{\theta}_{c}, \boldsymbol{\theta}_{0})$ 
be the parameter in a historical study, 
where $\boldsymbol{\theta}_{c}$ is a subset of  the common parameters. 
Now  
\begin{equation}
\pi(\boldsymbol{\theta},\delta|D_{0}) \propto 
\frac{ \big\{\int_{\boldsymbol{\Theta}_{0}}L(\boldsymbol{\theta}_{c}, \boldsymbol{\theta}_{0} |D_{0})^\delta
\pi_{0}(\boldsymbol{\theta}_{c}, \boldsymbol{\theta}_0) d \boldsymbol{\theta}_0 \big\}\pi_{0}(\boldsymbol{\theta}_{1})\pi_{0}(\delta)
 }{\int_{\boldsymbol{\Theta}_{c}} \big\{
 \int_{\boldsymbol{\Theta}_{0}}L(\boldsymbol{\theta}_{c}, \boldsymbol{\theta}_{0} |D_{0})^\delta
\pi_{0}(\boldsymbol{\theta}_{c}, \boldsymbol{\theta}_0) d \boldsymbol{\theta}_0 \big\} d\boldsymbol{\theta}_{c}} 
\label{eq:partialnpp}
\end{equation}
defines the partial borrowing normalized power prior, 
where $\boldsymbol{\Theta}_{0}$ and $\boldsymbol{\Theta}_{c}$ denote 
the parameter spaces of $\boldsymbol{\theta}_{0}$ and 
$\boldsymbol{\theta}_{c}$, respectively. 
In this case, the dimensions of $\boldsymbol{\Theta}_{0}$ and $\boldsymbol{\Theta}_{c}$ can be different, which is another 
advantage of using the prior in (\ref{eq:partialnpp}).

In addition, for model with latent variables $\boldsymbol{\xi}$, 
one can also extend the fixed borrowing 
to a random $\delta$ under the normalized power prior framework. 
Denote $g(\boldsymbol{\xi})$ the distribution of $\boldsymbol{\xi}$  
and assume $\boldsymbol{\theta}$ is the parameter of interest,  
we have two strategies to 
construct a power prior for $\boldsymbol{\theta}$ when $\delta$ is fixed. 
One way is to discount directly on the likelihood of $D_0$ 
expressed as $\int_{\boldsymbol{\Xi}} 
L(\boldsymbol{\theta} |D_0, \boldsymbol{\xi}) g(\boldsymbol{\xi}) 
d \boldsymbol{\xi}$, where 
$\boldsymbol{\Xi}$ denotes the domain of $\boldsymbol{\xi}$. 
The normalized power prior is of the form 
\begin{equation}
\pi(\boldsymbol{\theta},\delta|D_{0}) \propto 
\frac{ \left\{ \int_{\boldsymbol{\Xi}} 
L(\boldsymbol{\theta} |D_{0}, \boldsymbol{\xi})
g(\boldsymbol{\xi}) d \boldsymbol{\xi} \right\}^{\delta}
\pi_{0}(\boldsymbol{\theta})\pi_{0}(\delta)} 
{\int_{\boldsymbol{\Theta}}
 \left\{\int_{\boldsymbol{\Xi}}
 L(\boldsymbol{\theta}| D_{0}, \boldsymbol{\xi})
 g(\boldsymbol{\xi}) d\boldsymbol{\xi} \right\}^\delta 
\pi_{0}(\boldsymbol{\theta}) d\boldsymbol{\theta}}.
\label{eq:npplatent}
\end{equation}

Another borrowing strategy is to discount the likelihood 
of $D_0$ conditional on $\boldsymbol{\xi}$,  
while $g(\boldsymbol{\xi})$ is not discounted such that the 
power prior with $\delta$ fixed has the form 
$\pi_{0}(\boldsymbol{\theta}) \int_{\boldsymbol{\Xi}} 
L(\boldsymbol{\theta} |D_0, \boldsymbol{\xi})^{\delta} g(\boldsymbol{\xi}) 
d \boldsymbol{\xi}$. 
\cite{Ibrahim15} named such a prior {\it partial discounting power prior}.
We propose its counterpart beyond a fixed $\delta$, 
the {\it partial discounting normalized power prior}, which is formulated as
\begin{equation}
\pi(\boldsymbol{\theta},\delta|D_{0}) \propto 
\frac{ \left\{\int_{\boldsymbol{\Xi}}
L(\boldsymbol{\theta} |D_{0}, \boldsymbol{\xi})^\delta
g(\boldsymbol{\xi}) d \boldsymbol{\xi} \right\}
\pi_{0}(\boldsymbol{\theta})\pi_{0}(\delta)}
{\int_{\boldsymbol{\Theta}}
 \left\{\int_{\boldsymbol{\Xi}}
 L(\boldsymbol{\theta}| D_{0}, \boldsymbol{\xi})^\delta 
 g(\boldsymbol{\xi}) d \boldsymbol{\xi} \right\}
\pi_{0}(\boldsymbol{\theta}) d \boldsymbol{\theta}}.
\label{eq:discountnpp}
\end{equation}
\cite{Ibrahim15} argued that the partial discounting power prior is preferable 
due to both practical reasons and computational advantages. 
Both of the (\ref{eq:npplatent}) and (\ref{eq:discountnpp}) 
can be extended to models with random effects, 
in which the distribution $g(\boldsymbol{\xi})$ 
may depend on additional unknown variance parameters.  

Finally, we note that in the complex data analysis practice,
the extensions described above might be combined. For example, 
one can consider a partial borrowing normalized power prior with 
multiple historical data, where the borrowing is carried out only 
through some selected mutual parameters. 
Another example is in \cite{Chen14}, 
where the partial borrowing power prior 
is used in the presence of latent variables.  
Further variations for specific problems will be explored elsewhere.

\subsection{Computational Considerations in the Normalized Power Prior }
For the normalized power prior, 
the only computational effort in addition to that of the 
joint power prior is to calculate the scale factor 
$C(\delta) = \int_{\boldsymbol{\Theta}} L(\boldsymbol{\theta}|D_{0})^\delta 
\pi_{0}(\boldsymbol{\theta})
d\boldsymbol{\theta}$. 
In some models the integral can be calculated 
analytically up to a normalizing constant, so 
$\pi(\boldsymbol{\theta}, \delta|D_{0},D)$ 
can be expressed in closed forms. 
The posterior sample from   
$\pi(\boldsymbol{\boldsymbol{\theta}}, \delta|D_{0},D)$
can be obtained by first sampling from 
$\pi(\delta| \boldsymbol{\theta}, D_{0},D)$ or 
$\pi(\delta|D_{0},D)$, then from 
$\pi(\theta_{i}|\boldsymbol{\theta}_{-i}, \delta, D_{0},D)$, 
where $\boldsymbol{\theta}_{-i}$ is $\boldsymbol{\theta}$ 
without the $i^{th}$ element. 
It is typically achieved by using a Metropolis-Hastings algorithm 
\citep{Chib95} for $\delta$, followed by Gibbs sampling 
for each $\theta_i$. 

However, 
$C(\delta)$ needs to be calculated numerically in some models.  
General Monte Carlo methods to calculate the 
normalizing constant in the Bayesian computation 
can be applied.
Since the integrand includes a likelihood function powered to 
$\delta \in [0,1]$, 
we consider the following approach, 
which best tailored to the specific form of the integral. 
It is based on a variant of the algorithm in \cite{Friel08} and 
\cite{Rosmalen18} 
using the idea of path sampling \citep{GelmanMeng98}. 
The key observation is that 
$\log C(\delta)$ can be expressed as an integral of the 
expected log-likelihood of historical data, 
where the integral is calculated with respect to a bounded one-dimensional parameter. 
This identity can be written as
\begin{equation} \label{eq:calpha}
\log C(\delta) = \int_{0}^{\delta} 
E_{\pi(\boldsymbol{\theta} | D_0, \delta^* )} 
\{ \log [ L( \boldsymbol{\theta}|D_0 ) ] \} d {\delta^*},
\end{equation}
which is an adaptive version of the results from \cite{Friel08}. 
Proof is shown in \ref{appa}. 
For given $\delta^*$, 
the expectation in (\ref{eq:calpha}) is evaluated with respect to the density  
$\pi(\boldsymbol{\theta} | D_0, \delta^{*} ) \propto
L(\boldsymbol{\theta}|D_{0})^{\delta^{*}}\pi_{0}(\boldsymbol{\theta})$. 
Therefore the integrand can be calculated 
numerically if we can sample from  
$\pi(\boldsymbol{\theta} | D_0, \delta^{*})$. 
This is the prerequisite to implement the power prior 
with a fixed power parameter; 
hence no extra condition is required to calculate $\log C(\delta)$ 
using (\ref{eq:calpha}). 
By choosing an appropriate sequence of $\delta^{*}$ we can  
approximate the integral numerically. 

When sampling from the posterior 
$\pi(\boldsymbol{\theta}, \delta| D_0, D ) $ using the normalized power prior, 
 $C(\delta)$ needs to be calculated for every iteration. 
 \cite{Rosmalen18} suggested that the function 
$\log C(\delta)$ can be well approximated by 
linear interpolation.
Since $\delta$ is bounded, it is recommended to calculate  
a sufficiently large number of the $\log C(\delta)$ for 
different $\delta$ on a fine grid before the posterior sampling, 
then use a piecewise linear interpolation at each iteration during the posterior sampling.
In addition to the power prior with fixed $\delta$, 
the only computational cost is to determine $\log C(\delta)$ for 
selected values of $\delta \in [0, 1]$ as {\it knots}. 
Details of a sampling algorithm is 
provided in \ref{appb}.

Sampling from the density $\pi(\boldsymbol{\theta} | D_0, \delta^{*})$ 
can be computationally intensive in some models. 
Therefore the knots should be carefully selected given 
limited computational budget. A rule of thumb 
based on our empirical evidence is to select more grid points close to $0$, 
to account for the larger deviation from piecewise linearity in   
$\log C(\delta)$ when $\delta \to 0$. An example is to use  
$\left\{ \delta_{s} = (s/S)^{c} \right\}_{s = 0}^{S}$ with $c > 1$. 
Recently, \cite{Carvalho20} noted that $C(\delta)$ is 
a strictly convex function but not necessarily 
monotonic. They design primary 
grid points by prioritizing the region where the derivative 
$C'(\delta)$ is close to $0$, then use a generalized additive model to 
interpolate values on a larger grid. 
In practice, one may consider combining the two strategies above 
by adding some grid points used by \cite{Carvalho20} into the original design 
$\left\{ \delta_{s} = (s/S)^{c} \right\}_{s = 0}^{S}$. 
In addition, when $C(\delta)$ is not monotone, 
piecewise linear interpolation with limited number of grid points 
also needs to be cautious, especially around the region where $C'(\delta)$ change signs.


\subsection{Normalized Power Prior Approach for Exponential 
Family}\label{subsec:expfamily}

In this section we discuss how to make inference on  
parameter $\boldsymbol{\theta}$ (scalar or vector-valued) in an exponential
family, incorporating both the current data $D=(x_{1},\ldots,x_{n})$ 
and the historical data $D_{0}=(x_{01},\ldots,x_{0n_{0}})$. 
Suppose that the data comes from an exponential family with probability
density function or probability mass function of the form \citep{CasellaBerger01}
\begin{equation}\label{eq:expfamdensity}
f(x|\boldsymbol{\theta})=h(x) \exp \bigg\{\sum\limits_{i=1}^{k}
w_{i}(\boldsymbol{\theta})t_{i}(x) + \tau(\boldsymbol{\theta}) \bigg\},
\end{equation}
where the dimension of $\boldsymbol{\theta}$ is no larger than $k$. Here
$h(x)\geq 0$ and $t_{1}(x),\ldots,t_{k}(x)$ are real-valued functions
of the observation $x$, 
and $w_{1}(\boldsymbol{\theta}),\ldots,w_{k}(\boldsymbol{\theta})$ are
real-valued functions of the parameter $\boldsymbol{\theta}$. Define $
\underline{w}(\boldsymbol{\theta})=
\left(w_1(\boldsymbol{\theta}),\dots,w_k(\boldsymbol{\theta})\right)'$.
Furthermore, define 
\begin{align}\label{eq:compatibilitystat}
&
\underline{T}(\underline{x})
=\left(\frac{1}{n}\sum\limits_{j=1}^{n}
t_{1}(x_{j}),\ldots.,\frac{1}{n}\sum_{j=1}^{n}t_{k}(x_{j})\right)'
\end{align}
as the \emph{compatibility statistic} to measure how compatible a
sample $\underline{x}=(x_{1},\ldots.,x_{n})$ is with other samples in
providing information about $\boldsymbol{\theta}$. 
The density function of the current data can be expressed as
\begin{align}
\label{eq:currentdensity}
 f(D|\boldsymbol{\theta})=h(D)
 \exp\left\{n[\underline{T}(D)'\underline{w}
(\boldsymbol{\theta})+\tau(\boldsymbol{\theta})]\right\},
\end{align} 
where $h(D)=\prod_{j=1}^{n}h(x_{j})$ and $\underline{T}(D)$ stands for the
compatibility statistic related to the current data $D$.
Accordingly, the compatibility statistic and the density function
similar to (\ref{eq:compatibilitystat}) and (\ref{eq:currentdensity})
 for the historical data $D_0$ can be defined as well.
The joint posterior of $(\boldsymbol{\theta}, \delta)$ can be written as
\begin{equation}\label{eq:expfamjointpost}
\pi(\boldsymbol{\theta},\delta|D_{0},D)\propto
\frac{\exp\left\{ [\delta
n_0\underline{T}(D_0)'+n\underline{T}(D)']
\underline{w}(\boldsymbol{\theta})
+(\delta n_{0}+n) \tau(\boldsymbol{\theta}) \right\}
\pi_{0}(\boldsymbol{\theta})\pi_{0}(\delta)}
{\int_{\boldsymbol{\Theta}} \exp \left\{\delta n_{0}[
\underline{T}(D_0)'\underline{w}(\boldsymbol{\theta}) + 
\tau(\boldsymbol{\theta})] \right\} \pi_{0}(\boldsymbol{\theta}) 
 d\boldsymbol{\theta}}.
\end{equation}
Integrating $\boldsymbol{\theta}$ out from (\ref{eq:expfamjointpost}), 
the marginal posterior distribution of $\delta$ is given by
\begin{equation*}
\pi(\delta|D_{0},D) \propto \pi_{0}(\delta) \frac{\int_{\boldsymbol{\Theta}}
\exp\left\{[\delta
n_0\underline{T}(D_0)'+n\underline{T}(D)']\underline{w}(\boldsymbol{\theta})
+(\delta n_{0}+n) \tau(\boldsymbol{\theta}) \right\} \pi_{0}(\boldsymbol{\theta})
d\boldsymbol{\theta}}{\int_{\boldsymbol{\Theta}} \exp \left\{\delta n_{0}[
\underline{T}(D_0)'\underline{w}(\boldsymbol{\theta}) + 
\tau(\boldsymbol{\theta})] \right\} \pi_{0}(\boldsymbol{\theta})  d\boldsymbol{\theta}}.
\end{equation*}
The behavior of the power parameter $\delta$ can be examined from
this marginal posterior distribution. 
Similarly, the marginal
posterior distribution of $\boldsymbol{\theta}$ can be derived by integrating
$\delta$ out in $\pi(\boldsymbol{\theta},\delta|D_{0},D)$, 
but it often does not have a closed form. Instead the posterior distribution 
of $\boldsymbol{\theta}$ given $D_0$, 
$D$ and $\delta$ is often in a more familiar form.
Therefore we may learn the characteristic of the marginal posterior
of $\boldsymbol{\theta}$ by studying the conditional posterior distribution
$\pi(\boldsymbol{\theta}|D_{0},D,\delta)$, 
together with $\pi(\delta|D_{0},D)$. 

In the following subsections we provide three examples of the 
commonly used distributions, 
where the posterior marginal density (up to a normalizing constant) 
of $\delta$ can be expressed in closed forms. 
It can be extended to many other distributions as well 
by choosing appropriate initial priors 
$\pi_{0}(\boldsymbol{\theta})$.  

\subsubsection{\textbf{Bernoulli Population}}
Suppose we are interested in making inference on the probability
of success $p$ from a Bernoulli population with multiple replicates. 
Assume the total number of successes in the historical and the 
current data are $y_0 = \sum_{i=1}^{n_0} x_{0i}$ and 
$y=\sum_{i=1}^{n}x_i$ respectively, with the corresponding total number of trials $n_0$ and $n$. The
joint posterior distribution of $p$ and $\delta$ can be easily derived as the result below and the proof is omitted.

\smallskip
\noindent
\emph{\textbf{Result 1.}} Assume that the initial prior distribution
of $p$ follows a $\text{Beta}(\alpha,\beta)$ distribution, 
the joint posterior distribution of 
$(p,\delta)$ can be expressed as 
\begin{equation*}
\pi(p,\delta|D_{0},D) \propto \pi_{0}(\delta)
\frac{p^{\delta y_0+y{+\alpha-1}}
(1-p)^{\delta(n_0-y_0)+n-y+{\beta-1}}
 }{B(\delta
y_{0}+\alpha,\delta(n_{0}-y_{0})+\beta)},
\end{equation*}
where $B(a,b)=\frac{\Gamma(a)\Gamma(b)}{\Gamma(a+b)}$ stands for
the beta function.

Integrating $p$ out in $\pi(p,\delta|D_{0},D)$, the marginal
posterior distribution of $\delta$ can be expressed as
\begin{equation*}
\pi(\delta|D_{0},D) \propto \pi_{0}(\delta) \frac{B(\delta
y_{0}+y+\alpha,\delta(n_{0}-y_{0})+n-y+\beta)}{B(\delta
y_{0}+\alpha,\delta(n_{0}-y_{0})+\beta)}.
\end{equation*}
The conditional posterior distribution of $p$ given $\delta$ follows
a $\text{Beta} (\delta y_0+y+{\alpha},
\delta(n_0-y_0)+n-y+{\beta})$ distribution. 
However, the marginal posterior distribution of $p$ does not have a 
closed form.

\subsubsection{\textbf{Multinomial Population}}
\label{sec:multinomial}
As a generalization of the Bernoulli/binomial to $k\geq3$ categories, 
in a multinomial population assume we observe 
 historical data $D_0=(y_{01},y_{02},\ldots,y_{0k})$ and the current data $D=(y_{1},y_{2},\ldots,y_{k})$, 
 with each element represents the number of success 
 in that category. Let 
 $n_0=\sum_{i=1}^{k}y_{0i}$ and  $n=\sum_{i=1}^{k}y_{i}$. 
 Suppose the parameter of interest is 
 $\boldsymbol{\theta}=(\theta_1,\theta_2,\ldots,\theta_k)$ 
 which adds up to 1.  
 We have the following results below. 
 
\noindent
\emph{\textbf{Result 2.}} 
Assume the initial prior of $\boldsymbol{\theta}$ follows a 
Dirichlet distribution with $\pi_{0}(\boldsymbol{\theta})\sim 
\text{Dir}(\alpha_1,\alpha_2,\ldots,\alpha_k)$, the joint posterior of 
$(\boldsymbol{\theta}, \delta)$ can be expressed as 
 \begin{equation*}
  \pi(\boldsymbol{\theta},\delta | D_0, D)\propto 
  \pi_{0}(\delta)
  \prod_{i=1}^{k}\theta_i^{y_{0i}\delta+y_{i}+\alpha_i-1}
  \frac{\Gamma\left(n_{0}\delta + \sum_{i=1}^{k} \alpha_i \right)}
  {\prod_{i=1}^{k}\Gamma(y_{0i}\delta+\alpha_i)},
 \end{equation*}
  where $\Gamma(\cdot)$ stands for the gamma function.

The marginal posterior of $\delta$ can be derived by 
integrating $\boldsymbol{\theta}$ out as 
 \begin{equation*}
  \pi(\delta | D_0,D) \propto \pi_{0}(\delta)
  \frac{
  \Gamma\left(n_{0}\delta + \sum_{i=1}^{k} \alpha_i \right)
  \prod_{i=1}^{k} \Gamma(y_{0i}\delta+y_{i}+\alpha_i)
  }{\Gamma\left(n + n_{0}\delta + \sum_{i=1}^{k} \alpha_i \right)
  \prod_{i=1}^{k}\Gamma(y_{0i}\delta+\alpha_i)
  }.
 \end{equation*}
Similar to the Bernoulli case, the marginal posterior distribution 
of $\boldsymbol{\theta}$ does not have a closed form. 
The conditional posterior distribution of $\boldsymbol{\theta}$ given 
$\delta$ follows a Dirichlet distribution with 
$\text{Dir} (\delta y_{01} + y_1 + \alpha_1, 
\ldots, \delta y_{0k} + y_k + \alpha_k)$.

\subsubsection{\textbf{Normal Linear Model and Normal Population}}
\label{sec:LM}

Suppose we are interested in making inference on the regression parameters $\boldsymbol{\beta}$ from a linear model with current data \begin{eqnarray}\label{eq:LM}
\boldsymbol{Y}=\mathbf{X}\boldsymbol{\beta} +\boldsymbol{\epsilon},\text{ with }\boldsymbol{\epsilon}\sim \text{MVN}(\boldsymbol{0},\sigma^2 I_n),
\end{eqnarray} where the dimension of vector $\boldsymbol{Y}$ is $n$ and that of $\boldsymbol{\beta}$ is $k$. Similarly, we assume the historical data has the form $\boldsymbol{Y}_0 = \mathbf{X}_0\boldsymbol{\beta} +\boldsymbol{\epsilon}_0$, with $\boldsymbol{\epsilon}_0\sim \text{MVN}(\boldsymbol{0},\sigma^2 I_{n_0})$. Assume that both $\mathbf{X}_0'\mathbf{X}_0$ and $\mathbf{X}'\mathbf{X}$ are positive definite. Define 
\begin{align*}
    \hat{\boldsymbol{\beta}}_0&=(\mathbf{X}_{0}'\mathbf{X}_{0})^{-1}
    \mathbf{X}_{0}'\boldsymbol{Y}_{0}, ~~
    {S}_0 = (\boldsymbol{Y}_0-\mathbf{X}_{0} \hat{\boldsymbol{\beta}}_0)'(\boldsymbol{Y}_0-\mathbf{X}_{0} \hat{\boldsymbol{\beta}}_0),\\
        \hat{\boldsymbol{\beta}}&=(\mathbf{X'X})^{-1}\mathbf{X}'\boldsymbol{Y}, \text{  and }
    {S} = (\boldsymbol{Y}-\mathbf{X} \hat{\boldsymbol{\beta}})'(\boldsymbol{Y}-\mathbf{X} \hat{\boldsymbol{\beta}}).
\end{align*} 

Now, let's consider a conjugate initial prior for 
$(\boldsymbol{\beta},\sigma^2)$ as the following. 
$\pi_{0}(\sigma^2)\propto \sigma^{-2a}$, with $a>0$, and $\boldsymbol{\beta}|\sigma^2$ either has a MVN$(\boldsymbol{\mu}_0,\sigma^2 \boldsymbol{R}^{-1})$ distribution, which includes the Zellner's $g-$prior \citep{Zellner86} or $\pi_{0}(\boldsymbol{\beta}|\sigma^2)\propto 1$, 
which is a noninformative  prior. Here we assume $\boldsymbol{R}$ as a known positive definite matrix.
Hence, the initial prior can be written as
\begin{equation}\label{eq:ConjPrior}
    \pi_{0}(\boldsymbol{\beta},\sigma^2) \propto\frac{1}{(\sigma^2)^{a+\frac{kb}{2}}}
    \exp \left\{ {-\frac{b}{2\sigma^2}\left(\boldsymbol{\beta}-\boldsymbol{\mu}_0\right)'
    \boldsymbol{R}\left(\boldsymbol{\beta}-\boldsymbol{\mu}_0\right)}
    \right\},
    \text{ with }b=0\text{ or }1.
\end{equation}
We have the following theorem whose proof is given in \ref{appa}.

\begin{thm}\label{thm:LM}
With the set up above for the normal linear model (\ref{eq:LM}) 
and the initial prior of $(\boldsymbol{\beta},\sigma^2)$ as in (\ref{eq:ConjPrior}), 
suppose the initial prior of $\delta$ is 
$\pi_{0}(\delta)$. Then, the following results can be shown.
\begin{itemize}
    \item[(a)] The normalized power prior distribution of $(\boldsymbol{\beta},\sigma^2,\delta)$ is \begin{align*}
   \pi(\boldsymbol{\beta},\sigma^2,\delta |D_0)\propto \frac{\pi_{0}(\delta)M_0(\delta)}{\left(\sigma^2\right)^{\frac{\delta n_0+kb}{2}+a}}\exp\left\{-\frac{1}{2\sigma^2}\left[\delta\left\{{S}_0 + b{H}_0(\delta)\right\} +  {Q}(\delta,\boldsymbol{\beta})\right]\right\},     
    \end{align*} where 
    \begin{align*}
        {Q}(\delta,\boldsymbol{\beta}) &= (\boldsymbol{\beta}-\boldsymbol{\beta}^*)'\left(b\boldsymbol{R} +\delta\mathbf{X}_0'\mathbf{X}_0\right)(\boldsymbol{\beta}-\boldsymbol{\beta}^*),\\
        \boldsymbol{\beta}^* &=\left(b \boldsymbol{R} +\delta\mathbf{X}_0'\mathbf{X}_0\right)^{-1}\left
        (b\boldsymbol{R}\boldsymbol{\mu}_0 +\delta \mathbf{X}_0'\mathbf{X}_0\hat{\boldsymbol{\beta}}_0\right),\\
        {H}_0(\delta) &= \left(\boldsymbol{\mu}_0 -\hat{\boldsymbol{\beta}}_0\right)'\mathbf{X}_0'\mathbf{X}_0\left(b
        \boldsymbol{R} +\delta\mathbf{X}_0'\mathbf{X}_0\right)^{-1}
        \boldsymbol{R}\left(\boldsymbol{\mu}_0 -\hat{\boldsymbol{\beta}}_0\right),\text{ and }\\M_0(\delta) &= \frac{\left|b\boldsymbol{R}+\delta \mathbf{X}_0'\mathbf{X}_0\right|^{\frac{1}{2}}}{\Gamma\left(\frac{\delta n_0 + (b-1)k}{2}+a-1\right)}\left\{\delta\frac{{S}_0+b{H}_0(\delta)}{2}\right\}^{\frac{\delta n_0 + (b-1)k}{2}+a-1}.
    \end{align*} 

    \item[(b)] The marginal posterior density of $\delta$, given $(D_0,D)$, can be expressed as $$\pi(\delta|D_0,D)\propto \frac{\pi_{0}(\delta)\left|b\boldsymbol{R}+\delta \mathbf{X}_0'\mathbf{X}_0\right|^{\frac{1}{2}}
    \Gamma\left(\frac{n+\delta n_0+(b-1)k}{2}+a-1\right)}{\left|b\boldsymbol{R}+\delta \mathbf{X}_0'\mathbf{X}_0 +\mathbf{X}'\mathbf{X} \right|^{\frac{1}{2}} \Gamma\left(\frac{\delta n_0+(b-1)k}{2}+a-1\right)M(\delta)}, $$ 
    where $$M(\delta) = \left[\delta\left\{{S}_0 +b {H}_0(\delta)\right\}+ {S} +{H}(\delta)\right]^{\frac{n}{2}}\left[1+ 
    \frac{{S}+{H}(\delta)}{\delta\left\{{S}_0 +b {H}_0(\delta)\right\}}\right]^{\frac{\delta n_0+(b-1)k}{2}+a-1},$$
    $$\text{and }
    H(\delta) =  
    (\boldsymbol{\beta}^*-\hat{\boldsymbol{\beta}})'
    \mathbf{X}'\mathbf{X} \left(b\boldsymbol{R}
    +\delta\mathbf{X}_0'\mathbf{X}_0 
    + \mathbf{X}'\mathbf{X} \right)^{-1}
    \left(b\boldsymbol{R} +\delta\mathbf{X}_0'\mathbf{X}_0\right)
    (\boldsymbol{\beta}^*-\hat{\boldsymbol{\beta}}).
    $$

    \item[(c)] The conditional posterior distribution of $\boldsymbol{\beta}$, given $(\delta,D_0,D)$, is a multivariate Student {\em t}-distribution
    with location parameters $\boldsymbol{\mu}$, shape matrix $\mathbf{\Sigma}$, and the degrees of freedom $\nu$ as \begin{align*}
       \boldsymbol{\mu} &= \left(b\boldsymbol{R} +\delta\mathbf{X}_0'\mathbf{X}_0 
    + \mathbf{X}'\mathbf{X} \right)^{-1}\left\{ (b\boldsymbol{R}+\delta \mathbf{X}_0'\mathbf{X}_0)\boldsymbol{\beta}^* + \mathbf{X}'\mathbf{X}\hat{\boldsymbol{\beta}}\right\},       \\ \mathbf{\Sigma} &= \frac{{S}+{H}(\delta) + \delta \left\{{S}_0+b {H}_0(\delta)\right\}}{\nu}\left(b\boldsymbol{R} 
    +\delta \mathbf{X}_0'\mathbf{X}_0+ \mathbf{X}'\mathbf{X}\right)^{-1},\text{ and} \\\nu &= (b-1)k+\delta n_0+n+2a-2.
    \end{align*}
    \item[(d)] The conditional posterior distribution of $\sigma^2$, given $(\delta,D_0,D)$, follows an inverse-gamma distribution with shape parameter $\frac{(b-1)k+\delta n_0 +n}{2}+a-1$, and scale parameter $\frac{1}{2}\left[{S}+{H}(\delta) + \delta \left\{{S}_0+b {H}_0(\delta)\right\}\right]$.
\end{itemize} 
\end{thm} 

Theorem \ref{thm:LM} provides a general case for the normal linear model 
with certain conjugate prior structure. 
We can easily obtain the results for a regular normal population with such conjugate structure.  One of the results for a normal population $N(\mu,\sigma^2)$ with $\pi_{0}(\mu,\sigma^2)\propto \sigma^{-2a}$ and $\pi_{0}(\delta)\sim\text{Beta}(\alpha_{\delta},\beta_{\delta})$ can be found in \cite{Duan06a}.

\section{Optimality Properties of the Normalized Power
Prior}\label{sec:optimal}

In investigating the optimality properties of the normalized power priors, we use the idea of minimizing the weighted
Kullback-Leibler (KL) divergence \citep{Kullback51} that is similar to, but not the same as in \cite{Ibrahim03}. 

Recall the definition of the KL divergence,
\begin{equation*}
K(g,f)=\int_{\boldsymbol{\Theta}}
\log\bigg(\frac{g(\boldsymbol{\theta})}{f(\boldsymbol{\theta})}\bigg)
g(\boldsymbol{\theta}) d \boldsymbol{\theta},
\end{equation*}
where $g$ and $f$ are two densities with respect to Lebesgue
measure.  In \cite{Ibrahim03}, a loss function related to a target density $g$, denoted by $K_g$, is defined as the 
convex sum of the KL divergence 
between $g$ and two posterior densities. 
One is the posterior density 
without using any historical data, 
denoted by 
$f_0\propto L(\boldsymbol{\theta}|D)
\pi_{0}(\boldsymbol{\theta})$, 
and the other is the posterior 
density with the historical and
current data equally weighted, 
denoted by 
$f_1\propto L(\boldsymbol{\theta}|D_{0})L(\boldsymbol{\theta}|D)
\pi_{0}(\boldsymbol{\theta})$. The loss is defined as $$K_g = (1-\delta)K(g,f_0) +\delta K(g,f_1),$$ where the weight for $f_1$ is $\delta$.  
It is showed  that, when $\delta$ is given, 
the unique minimizer of $K_g$ is the posterior distribution derived using the 
power prior, i.e.,
\begin{equation*}\label{eq:conditionalpos}
    \pi(\boldsymbol{\theta}|D_{0},D,\delta) 
    \propto L(\boldsymbol{\theta}|D_{0})^\delta L(\boldsymbol{\theta}|D)
    \pi_{0}(\boldsymbol{\theta}).
\end{equation*} 
Furthermore, \cite{Ibrahim03} claim that the posterior derived from the joint power prior also minimizes $E_{\pi_{0}(\delta)}\left(K_g\right)$ when $\delta$ is random.

We look into the problem from a different angle. Since the prior for $\boldsymbol{\theta}$ without the historical data is $\pi_{0}(\boldsymbol{\theta})$ with $\int_{\boldsymbol{\Theta}}\pi_0(\boldsymbol{\theta}) d\boldsymbol{\theta} = 1$, 
we further denote the prior for $\boldsymbol{\theta}$ when fully utilizing the historical data as $\pi_1(\boldsymbol{\theta}) \propto \pi_{0}(\boldsymbol{\theta}) L(\boldsymbol{\theta}|D_0)$, with $\int_{\boldsymbol{\Theta}}\pi_1(\boldsymbol{\theta}) d\boldsymbol{\theta} = 1$. Clearly \begin{eqnarray}\label{eq:QD0}
\pi_1(\boldsymbol{\theta}) = Q(D_0)\pi_{0}(\boldsymbol{\theta}) L(\boldsymbol{\theta}|D_0),
\end{eqnarray} 
where $Q^{-1}(D_0) = \int_{\boldsymbol{\Theta}}\pi_{0}(\boldsymbol{\theta}) L(\boldsymbol{\theta}|D_0) d\boldsymbol{\theta}$ is a normalizing constant.

Suppose we have a prior $\pi_{0}(\delta)$. For any function $g(\boldsymbol{\theta}|\delta)$, define the 
expected weighted KL divergence between $g$ and $\pi_0$, and between $g$ and $\pi_1$ as \begin{eqnarray}\label{def:WeightedKL}
L_g = E_{\pi_{0}(\delta)}\left\{(1-\delta) K(g,\pi_0) + \delta K(g,\pi_1)\right\},
\end{eqnarray} where $0\le \delta\le 1$. We have the following theorem whose proof is given in \ref{appa}.

\begin{thm}\label{th:optimality}
\renewcommand{\baselinestretch}{1.7}
Suppose $\pi(\delta|D_0)=\pi_{0}(\delta)$. The function $g(\boldsymbol{\theta}|\delta,D_0)$ that minimizes the expected weighted KL divergence defined in (\ref{def:WeightedKL}) is 
\begin{equation*}
\pi^{\ast}(\boldsymbol{\theta}|\delta,D_0) = \frac{L(\boldsymbol{\theta}|D_{0})^\delta 
\pi_{0}(\boldsymbol{\theta}) }{\int_{\boldsymbol{\Theta}} 
L(\boldsymbol{\theta}|D_{0})^\delta \pi_{0}(\boldsymbol{\theta}) 
d\boldsymbol{\theta}},
\end{equation*} from which we deduce the normalized power prior $\pi(\boldsymbol{\theta},\delta|D_0)$ in (\ref{eq:powerprior}).
\end{thm}
Note that the last claim in Theorem \ref{th:optimality} comes from $$\pi(\boldsymbol{\theta},\delta|D_0) = \pi(\boldsymbol{\theta}|\delta,D_0) \pi(\delta|D_0) = \pi(\boldsymbol{\theta}|\delta,D_0) \pi_{0}(\delta).$$
The assumption of $\pi(\delta|D_0) =  \pi_{0}(\delta)$ indicates that the original prior of $\delta$ does not depend on $D_0$, which is reasonable.

\section{Posterior Behavior of the Normalized Power Prior
}\label{sec:BehaviorComp}

In this section we investigate the posteriors of both $\boldsymbol{\theta}$ and $\delta$ under different 
settings of the observed statistics. 
We show that by using the normalized power prior, 
the resulting posteriors can respond to the 
compatibility between $D_0$ and $D$ in an expected way. 
However, the posteriors are sensitive to different forms of the 
likelihoods under same data and model using the joint power 
priors.

\subsection{Results on the Marginal Posterior Mode of the Power 
Parameter}\label{subsec:ModeDelta}

Some theoretical results regarding the relationship between the 
posterior mode of $\delta$ and the \emph{compatibility statistic} defined in
(\ref{eq:compatibilitystat}) are given as follows. Their proofs are given in \ref{appa}.

\begin{thm}
\label{th:mode}
Suppose that historical data $D_{0}$ and current data $D$ are two
independent random samples from an exponential family given in
(\ref{eq:expfamdensity}). The compatibility statistic for 
$D_0$ and $D$ are
$\underline{T}(D_{0})$ and $\underline{T}(D)$ respectively 
as defined in (\ref{eq:compatibilitystat}).
Then the marginal posterior mode of $\delta$ is always $1$ under the
normalized power prior approach, if 
\begin{equation}
\label{eq:Mode1Cond}\frac{d}{d\delta}\log\pi_{0}(\delta)+h_1(D_0,D,\delta)
+n_0[\underline{T}(D_0)-\underline{T}(D)]'\underline{h}_2(D_0,
D,\delta)\ge 0,
\end{equation} for all $0\le\delta\le 1$, where
\begin{equation*}
  h_1(D_0,D,\delta)=\frac{n_0}{n}\int_{\boldsymbol{\Theta}}\log L(\boldsymbol{\theta}|D)
  [\pi(\boldsymbol{\theta}|D_0, D,\delta)-\pi(\boldsymbol{\theta}|D_0,\delta)]
  d\boldsymbol{\theta},
\end{equation*} and
\begin{equation*}
  \underline{h}_2(D_0,D,\delta)=\int_{\boldsymbol{\Theta}}\underline{w}
  (\boldsymbol{\theta}) [\pi(\boldsymbol{\theta}|D_0,
  D,\delta)-\pi(\boldsymbol{\theta}|D_0,\delta)]d\boldsymbol{\theta}.
\end{equation*}
\end{thm}

The first term in (\ref{eq:Mode1Cond}) is always non-negative if the
prior of $\delta$ is a nondecreasing function.  Hence, if one uses uniform prior on $\delta$, this term is zero. The second
term, $h_1(D_0,D,\delta)$, is always non-negative by using the
property of KL divergence. 
It is 0 if and only if $\pi(\boldsymbol{\theta}|D_0,
D,\delta)=\pi(\boldsymbol{\theta}|D_0,\delta)$, which means 
given $\delta$ and $D_0$, current data $D$ 
does not contribute to any information for $\boldsymbol{\theta}$. 
This could be a rare case. The third term in (\ref{eq:Mode1Cond})
depends on how close 
$\underline{T}(D_0)$ and $\underline{T}(D)$ are to each other.  
When $\underline{T}(D_0) = \underline{T}(D)$, the third term is zero, and hence
the posterior mode of $\delta$ is 1. 
Since $h_1(D_0,D,\delta)$ is non-negative, 
the posterior mode of $\delta$ may also achieve $1$ 
as long as the difference between $\underline{T}(D_{0})$ and
$\underline{T}(D)$ is negligible from a practical point of view. 
On the other hand, for the joint power prior approach, we 
have the following result.

\begin{thm}
\label{th:Mode0Result}
Suppose that current data $D$ comes from a population with density
function $f(x|\boldsymbol{\theta})$, and $D_{0}$ is a related historical data. 
Furthermore, suppose that the initial prior $\pi_{0}(\delta)$ is a
non-increasing function and the conditional posterior distribution
of $\boldsymbol{\theta}$ given $\delta$ is proper for any $\delta$. 
Then for any
$D_{0}$ and $D$, if \begin{equation}   \label{eq:Mode0Cond}
\underset{0\le\delta\le 1}{\max}\frac{\int
\pi_{0}(\boldsymbol{\theta})f(D|\boldsymbol{\theta})f(D_{0}|\boldsymbol{\theta})^\delta \log
f(D_{0}|\boldsymbol{\theta})d\boldsymbol{\theta}}{\int
\pi_{0}(\boldsymbol{\theta})f(D|\boldsymbol{\theta})
f(D_{0}|\boldsymbol{\boldsymbol{\theta}})^\delta d\boldsymbol{\theta}}<\infty,
\end{equation} then there exists at least one
positive constant $k_{0}$ such that $\pi(\delta|D_{0},D)$ has mode
at $\delta=0$ under the joint power prior,
where $L(\boldsymbol{\theta}|x)=k_{0} f(x|\boldsymbol{\theta})$.
\end{thm} 

The assumption in (\ref{eq:Mode0Cond}) is valid in the case that all
the integrals are finite positive values when $\delta$ 
is either 0 or 1.  
Usually this condition satisfies when $\pi_{0}(\boldsymbol{\theta})$ 
is smooth. The proof of this 
result is also given in the \ref{appa}.  
For a normal or a Bernoulli
population, our research reveals that $\pi(\delta|D_{0},D)$ has mode
at $\delta=0$ in many scenarios regardless of 
the level of compatibility between $D$ and $D_0$. 
Note that the results in Theorem \ref{th:Mode0Result} is not limited to exponential family distributions.

A primary objective of considering $\delta$ as random is to let the 
posterior inform the compatibility between the historical and 
the current data, given a vague initial prior on $\delta$. 
This allows adaptive borrowing according to the prior-data conflict. 
Theorem \ref{th:mode} indicates that, when the uniform initial prior of $\delta$ is 
used, the posterior of $\delta$ could potentially suggest 
borrowing more information from $D_0$ as long as $D$ is 
compatible with $D_0$. In practice, this has the potential to reduce 
the sample size required in $D$ in the design stage, and to provide 
estimates with high precision in the analysis stage. 
Theorem \ref{th:Mode0Result} shows that, on the other hand, 
if one considers the joint power prior 
with an arbitrary likelihood form and a smooth initial prior 
$\pi_{0}(\boldsymbol{\theta})$, it is possible that the posterior of 
$\delta$ could not inform the data compatibility. 
This suggests the opposite, 
meaning that adaptive borrowing might not be true when using the 
joint power prior; see Section \ref{subsec:DeltaSim} for 
more details. 



\subsection{
Posteriors of Model Parameters}\label{subsec:DeltaSim}
We investigate the posteriors of all model parameters  
in Bernoulli and normal populations, to illustrate that different forms of 
the likelihoods  
could result in different posteriors, which affects the borrowing strength. 

\begin{figure}[ht!]
\begin{center}
\psfig{figure=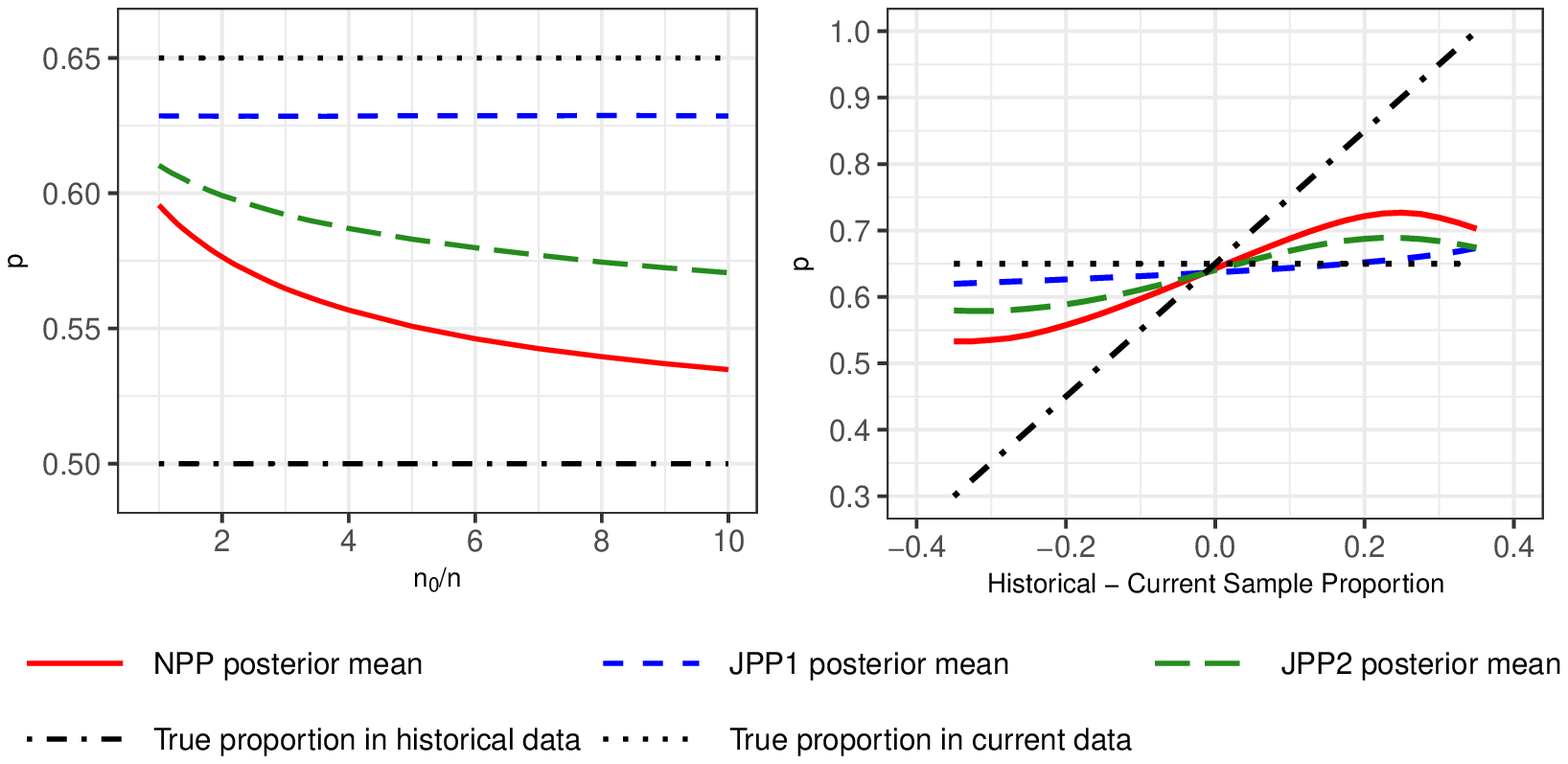,width=3.5in,height=1.8in} 
\\ \vspace{0.1 cm}
\psfig{figure=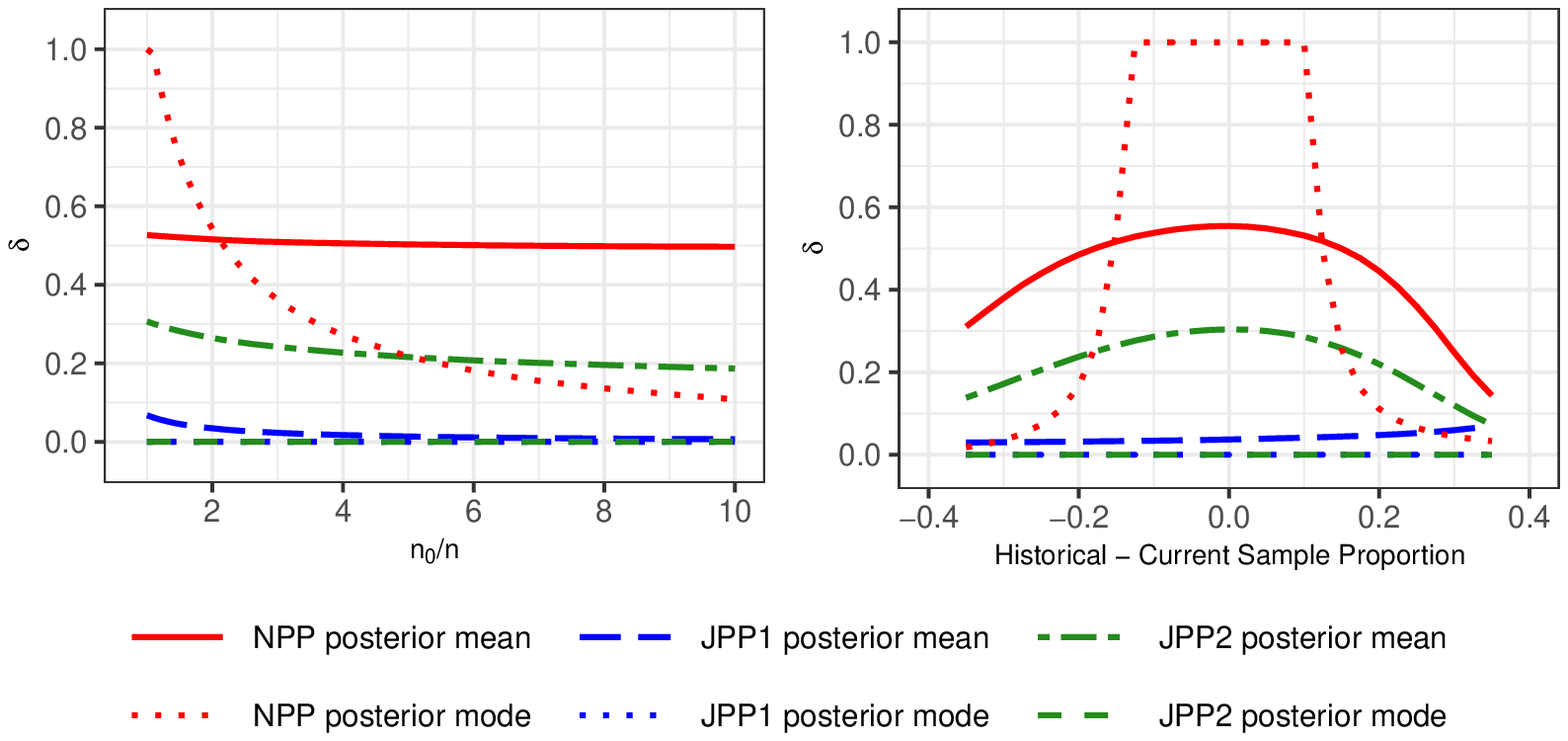,width=3.5in,height=1.8in} 
\end{center}
\caption{Posterior behavior of $p$ (top) and 
$\delta$ (bottom) for Bernoulli population 
		when $n=20$, $\hat{p}=0.65$. 
	    Left: $\hat{p}_{0}=0.5$ fixed and $n_0$ varies. Right: 
        $n_0=40$ fixed and $\hat{p}_0$ varies.}
\label{fi:ber-p-delta}
\end{figure}

For independent Bernoulli trials, 
two different forms of the likelihood functions are commonly used. 
One is based on the product of independent Bernoulli densities such that  
$L_{J1}(p|D_0)= p^{y_0} (1-p)^{n_0-y_0}$, and another is based on the 
sufficient statistic, the summation of the 
binary outcomes, which follows a binomial distribution
$L_{J2}(p|D_0)= c_1 p^{y_0} (1-p)^{n_0-y_0}$, where 
$c_1 = {\displaystyle {\binom {n_0}{y_0}}}$. 
Assuming $\pi_{0}(p) \sim \text{Beta}(\alpha, \beta)$, the  
corresponding posteriors are 
$$
\pi_{J1}(p, \delta|D_0, D) \propto \pi_{0}(\delta) 
p^{\delta y_0+y{+\alpha-1}}
(1-p)^{\delta(n_0-y_0)+n-y+{\beta-1}}
$$
and 
$$
\pi_{J2}(p, \delta|D_0, D) \propto 
c_{1}^{\delta} \pi_{J1}(p, \delta|D_0, D), 
$$
respectively. After marginalization we have 
\begin{align*}
\pi_{J1}(\delta|D_0, D) &\propto \pi_{0}(\delta) 
B(\delta y_{0}+y+\alpha,\delta(n_{0}-y_{0})+n-y+\beta), \\
\pi_{J2}(\delta|D_0, D) &\propto 
c_{1}^{\delta} \pi_{J1}(\delta|D_0, D). 
\end{align*}
We denote these two scenarios as JPP1 and JPP2 in Figure \ref{fi:ber-p-delta}. 

For the normal population, we also consider two different forms of the likelihood 
functions. 
One uses the product of $n_0$ independent 
normal densities 
$$
L_{J1}(\mu,\sigma^2|D_0) = (2\pi \sigma^2)^{-\frac{n_0}{2}} 
\exp \left\{-\frac{\sum_{i = 1}^{n_0}(x_{0i}-\mu)^2}{2\sigma^2} \right\}, 
$$
where 
${x}_{0i}$ is the value of the $i^{th}$ 
observation in $D_0$. 
Another less frequently used form is the density of sufficient statistics 
$f(\bar{x}_{0},s_0^{2}|\mu,\sigma^2)$, where $\bar{x}_{0}$ and $s_0^{2}$ are the sample mean and variance of $D_0$, respectively.
Since $\bar{x}_{0} \sim N\big(\mu, \frac{\sigma^2}{n_0} \big)$ and 
$\frac{(n_0-1)s_{0}^2}{\sigma^2}\sim {\chi}^2_{n_0-1}$, so 
$s_{0}^2 \sim \text{Gamma}\big(\frac{n_0-1}{2}, \frac{2\sigma^2}{n_0-1}\big)$ 
under the shape-scale parameterization. Then 
$$
L_{J2}(\mu,\sigma^2|D_0)  = 
c_{2}(\sigma^2)^{-\frac{n_0}{2}}
\exp\left\{-\frac{n_0(\bar{x}_{0} - \mu)^2 + (n_0-1)s_{0}^2}{2\sigma^2}\right\}, 
$$
where $\log c_{2} = 
(n_0 -3)\log s_0 + \frac{n_0 - 1}{2} \log\left(\frac{n_0 - 1}{2}\right) + 
\frac{1}{2} \log n_0 - \frac{1}{2} \log(2\pi) - \log \Gamma(\frac{n_0 - 1}{2})$. 

Similar to the Bernoulli case, we can easily derive their joint power
priors and the corresponding posteriors denoted as JPP1 and JPP2.
As a result, their log posteriors are differed by 
$-\frac{n_0 \delta}{2} \log(2\pi) - \delta \log(c_2)$. 
In the numerical experiment we use a $\text{Beta}(1,1)$ as the initial prior for $\delta$, and 
the reference prior $\pi_{0}(\mu,\sigma^2)\propto 1/\sigma^2$ 
\citep{BergerBernardo92} as the initial prior for
$(\mu,\sigma^2)$.

\begin{figure}[ht!]
\begin{center}
		\psfig{figure=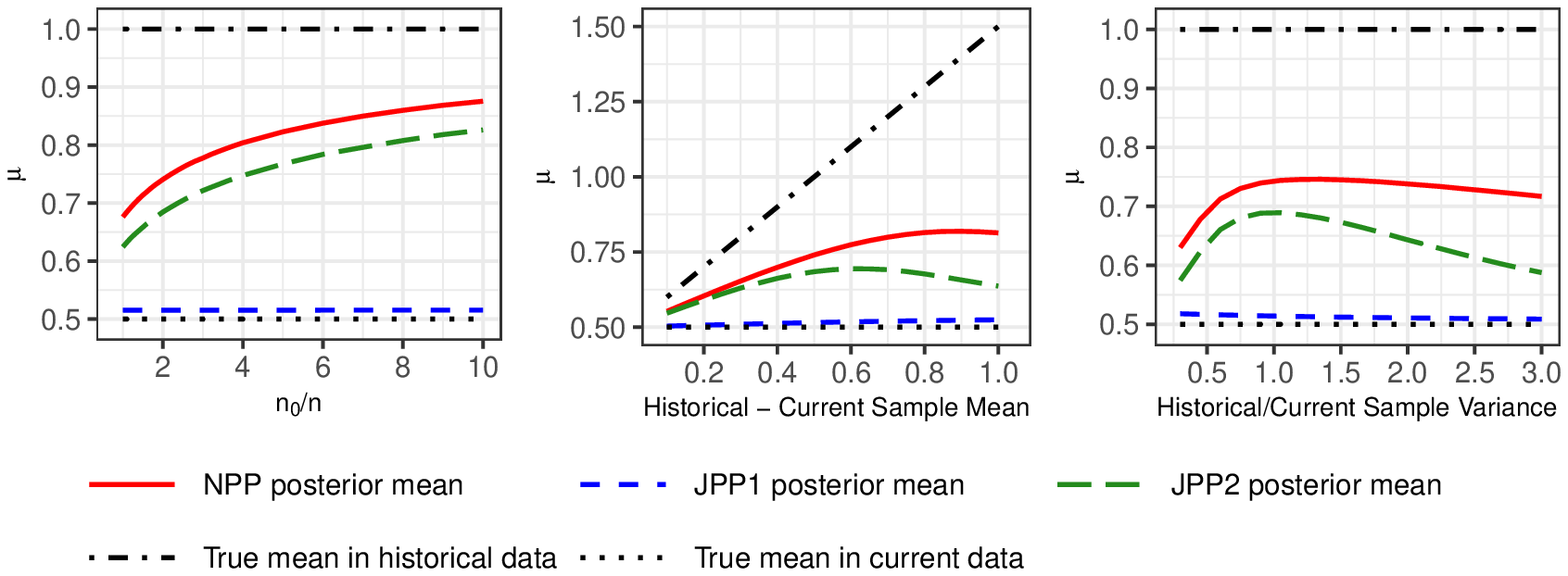,width=1.75
		in,height=5.2in,angle=-90}
		\\ \vspace{0.1 cm}
		\psfig{figure=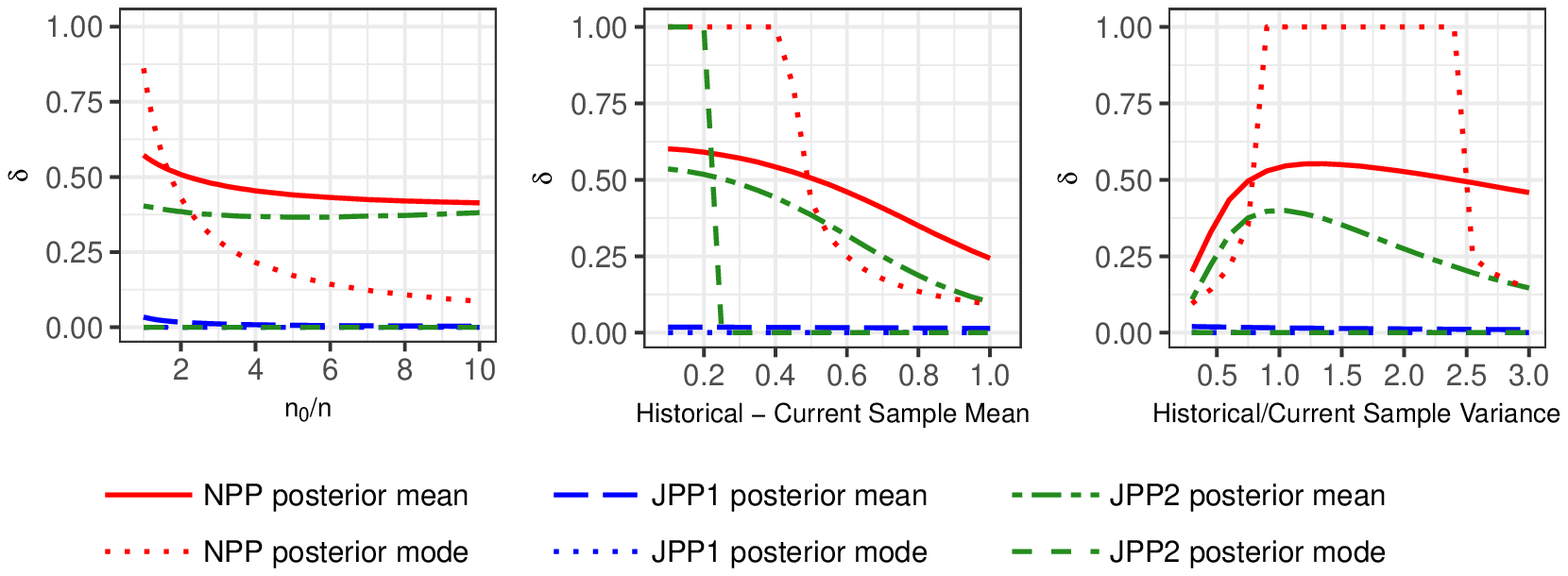,width=1.75
		in,height=5.2in,angle=-90}
\end{center}
		\caption{Posterior behavior of $\mu$ (top) and 
		$\delta$ (bottom) for normal population 
			when $n=20$, $\bar{x}=0.5$, $\hat{\sigma}^{2} = 1$. 
			Left: $\bar{x}_{0}=1$ and $\hat{\sigma}^{2}_{0} = 0.8$ fixed, $n_0$ varies. 
			Middle: $n_0=40$ and $\hat{\sigma}^{2}_{0} = 0.8$ fixed, $\bar{x}_{0}$ varies. 
			Right: $n_0=40$ and $\bar{x}_{0}=1$ fixed, $\hat{\sigma}^{2}_{0}$ varies. }\label{fi:normal-mu-delta}
\end{figure}

Figure \ref{fi:ber-p-delta} shows how the posteriors  
of $p$ and $\delta$ change with $n_0/n$ and $\hat{p}_0-\hat{p}$ 
in data simulated from the Bernoulli population, 
in which a $\text{Beta}(1,1)$ is used as the initial prior for both $p$ and $\delta$.
Figure \ref{fi:normal-mu-delta} 
shows how the posterior 
of $\mu$ and $\delta$ change with $n_0/n$, $\hat{\mu}_0-\hat{\mu}$ (for fixed $\hat{\sigma}^2_0$ and $\hat{\sigma}^2$), 
and  $\hat{\sigma}^2_0/\hat{\sigma}^2$ 
(for fixed $\hat{\mu}_0$ and $\hat{\mu}$) in the normal population. 

From both Figures \ref{fi:ber-p-delta} and \ref{fi:normal-mu-delta}, 
we observe, under the normalized power prior, 
the posterior mean of the parameter of interest 
($p$ in the Bernoulli population and 
$\mu$ in the normal population) 
are sensitive to the change of compatibility between $D$ and $D_0$. 
As the difference between the observed sample 
average of $D_0$ and $D$ 
increases, 
the posterior mean of both $p$ and $\mu$ are getting closer to the parameter estimate based on $D_0$ at the beginning, 
then going back to the parameter estimate based on $D$. 
For increasing $n_{0}/n$, the posterior mean 
are getting closer to the parameter estimate based on $D_0$.
Both of the posterior mean and mode of $\delta$ 
respond to the compatibility between $D_0$ and $D$ as expected. 
In addition, when the two samples
are not perfectly homogeneous, 
the posterior mode of $\delta$ can still 
attain $1$. 
This is reasonable because the historical 
population is subjectively believed to have similarity with the
current population with a modest amount of heterogeneous. 
These findings imply that the power parameter 
$\delta$ responds to data in a sensible way in the normalized power prior approach.

When using the joint power prior approach,  
we observe that the 
posteriors of the parameters $p$, $\mu$ and $\delta$ behave differently with   
different forms of the likelihoods.  
Despite a violation of the likelihood principle, 
the joint power prior might provide moderate adaptive borrowing 
under certain form of the likelihood. 
The degree of the adaptive borrowing is  
less than using the normalized power prior. 
Under another likelihood form in our illustration, 
the posteriors suggest almost no borrowing, 
regardless of how compatible these two samples are.

\section{Behavior of the Square Root of Mean Square Error 
under the Normalized Power Prior}
\label{sec:mse}

We now investigate the influence of borrowing historical data 
in parameter estimation using the square root of the mean square error 
(rMSE) as the criteria. Several different approaches are compared, 
including the full borrowing (pooling), no borrowing, 
normalized power prior, and joint power prior. 
Two different likelihood forms are used for $D_0$ in 
the joint power priors, with the same notation as in Section \ref{sec:BehaviorComp}. 
The rMSE obtained by the Monte Carlo method, defined as 
$\sqrt{\frac{1}{m} \sum\limits_{i=1}^{m}
(\hat{\boldsymbol{\theta}}^{(i)}-\boldsymbol{\theta})^{2}}$, 
is used for comparison, where $m$ is the number of 
Monte Carlo samples, $\boldsymbol{\theta}$ is the 
true parameter and $\hat{\boldsymbol{\theta}}^{(i)}$ is the estimate in the $i^{th}$ sample. We choose $m=5000$ in all  experiments. 

\subsection{Bernoulli Population} 
We first compute the rMSE of 
estimated $p$ in independent Bernoulli trials, 
where $p$ is the probability of success in 
the current population. 
Suppose the current  
data comes from a binomial($n$, $p$) distribution and the historical
data comes from a binomial($n_0$, $p_{0}$) distribution, with both $p$
and $p_{0}$ unknown. 
The posterior mean of $p$ is used as the estimate. 
In the simulation experiment we choose $n=30$, $p = 0.2$ or $0.5$, 
and $n_0 = 15, 30$ or $60$. 
We use the $\text{Beta}(1,1)$ 
as the initial prior for both $p$ and $\delta$.  

\begin{figure}[ht!]
\begin{center}
	\psfig{figure=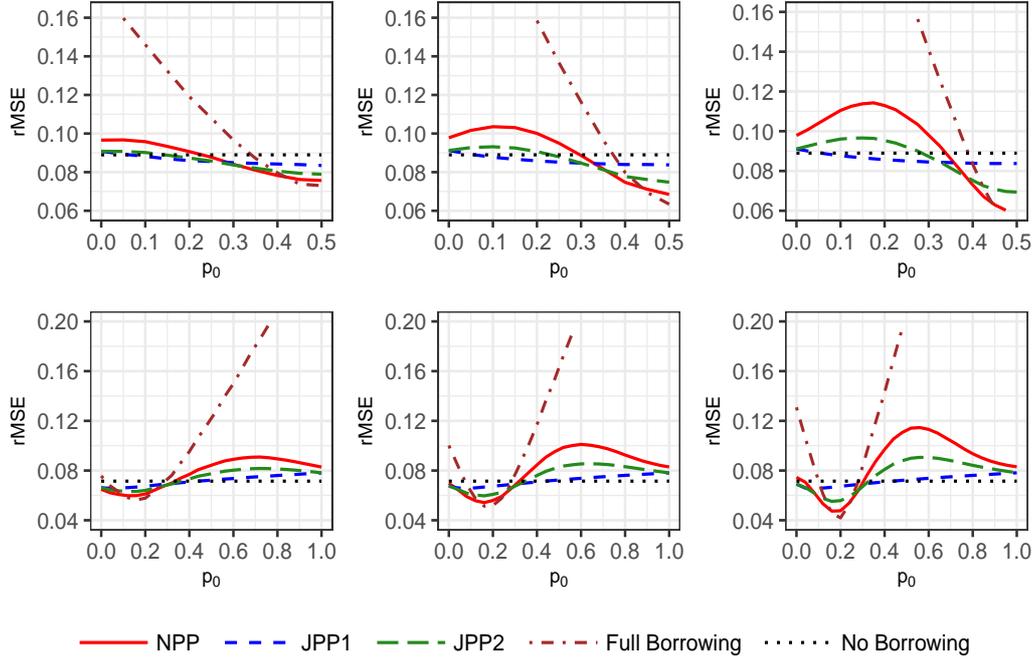,width=3.6 in,height=5.4in,angle=-90} 
\end{center}
	\caption{Square root of the MSE of $\hat{p}$ when 
	$n = 30$. Top: $p = 0.5$; Bottom: $p = 0.2$. 
	Left: $n_0 = 15$; Middle: $n_0 = 30$; Right: $n_0 = 60$.
	}\label{fi:mseber}
\end{figure}

Based on the results in Figure \ref{fi:mseber}, 
the normalized power prior approach yields the rMSE comparable to the 
full borrowing when the divergence between the current and the historical 
population is small or mild. 
As $|p-p_0|$ increases from $0$, 
both the posterior mean and the mode of 
$\delta$ will decrease on average. 
The rMSE of the posterior mean of $p$ will increase 
with $|p-p_0|$ when $p_0$ is near $p$. 
As the $|p_0-p|$ further increases, the posterior mean and 
mode of $\delta$ will automatically drop toward $0$ 
(Figure \ref{fi:deltaformse}), 
so the rMSE will then decrease and 
eventually drop to the 
level comparable to no borrowing. 
Also, when $|p-p_0|$ is small, the rMSE will decrease 
as $n_0$ increase, which implies when the divergence 
between the current and the historical populations is mild,
incorporating more historical data would result in 
better estimates using the normalized power prior. 
However, when $|p-p_0|$ is large, the 
rMSE will increase with $n_0$ in most scenarios. 
All plots from Figures \ref{fi:mseber} and \ref{fi:deltaformse} 
indicate that the normalized power prior approach provides adaptive borrowing.

For the joint power prior approaches, 
the prior with the likelihood expressed as the 
product of independent Bernoulli densities is similar to no 
borrowing while using the prior  
based on a binomial likelihood tends to 
provide some adaptive borrowing, 
with less information incorporated than using the normalized power prior. 
This is consistent with what we observed regarding their posteriors in 
Section \ref{sec:BehaviorComp}. 

\subsection{Normal Population}
We also investigate the rMSE of estimated $\mu$ in a normal 
population with unknown variance. 
Suppose that the current and historical samples  
are from normal $N(\mu,\sigma^2)$ 
and $N(\mu_{0},\sigma_{0}^2)$ populations respectively,
with both mean and variance unknown. 
Furthermore, the population mean $\mu$ is the parameter of interest, and the 
posterior mean is used as the estimate of $\mu$.

It can be shown that the marginal posterior distribution of 
$\delta$ only depends on $n_0$, $n_0/n$, $\sigma_{0}/\sigma$, 
and $(\mu_{0}-\mu)/\sigma$, and so does the rMSE.
Therefore we design two simulation settings,  
with $n=30$, $\mu=0$, $\sigma=1$, and $n_0 = 15, 30$ or $60$ under both settings.
In the first experiment we fix $\sigma_{0} = 1$,  
the heterogeneity is reflected 
by varying $\mu_0$ and therefore $(\mu_{0}-\mu)/\sigma$.
In the second experiment, we fix $\mu_0= 0.2$ so 
$(\mu_{0}-\mu)/\sigma $ is fixed at $0.2$. 
We change $\sigma_0$ at various levels resulting in 
changes in $\sigma_{0}/\sigma$.

\begin{figure}[ht!]
\begin{center}
	\psfig{figure=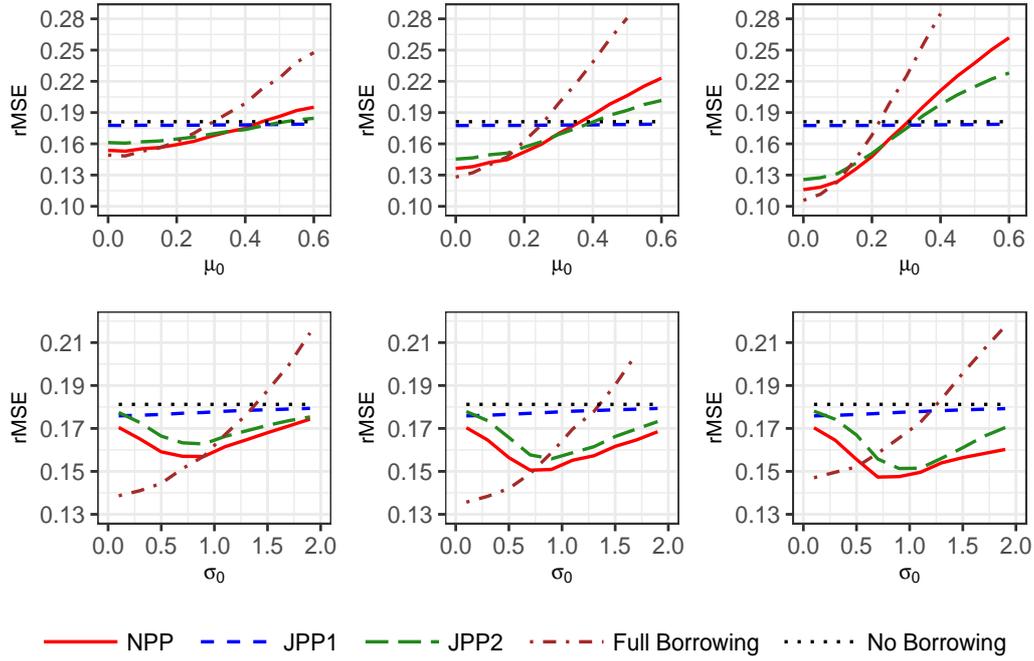,width=3.6 in,height=5.4in,angle=-90} 
\end{center}
	\caption{Square root of the MSE of $\hat{\mu}$ when 
	$n = 30$, $\mu = 0$, $\sigma = 1$. 
	Top: $\sigma_0 =  1$. Bottom: $\mu_0 = 0.2$. 
	Left: $n_0 = 15$; Middle: $n_0 = 30$; Right: $n_0 = 60$.
	}\label{fi:msenormal}
\end{figure}

\begin{figure}[ht!]
\begin{center}
	\psfig{figure=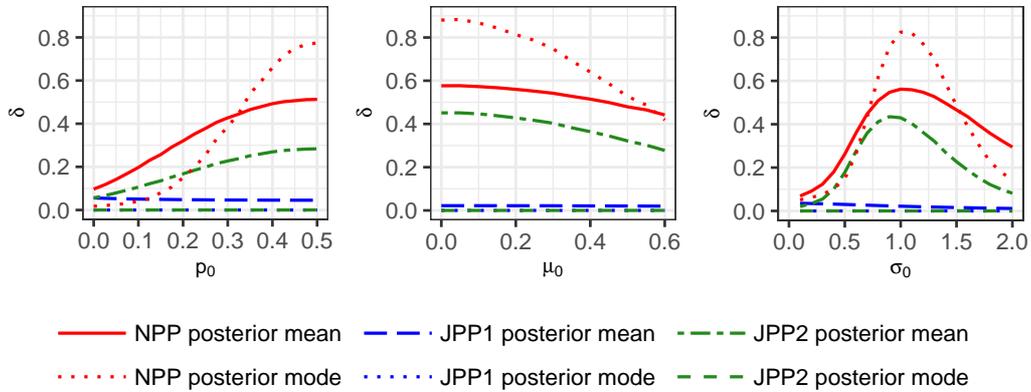,width=2.2 in,height=5.4in,angle=-90} 
\end{center}
	\caption{Average value of the posterior mean for  
	$\delta$ in simulated data with 
	$n = n_0 = 30$.  
	Left: Bernoulli population with $p=0.5$; 
	Middle: Normal population with $\mu=0$ and $\sigma = \sigma_0 = 1$; 
	Right: Normal population with $\mu=0$, $\mu_0=0.2$ and $\sigma = 1$. 
	}\label{fi:deltaformse}
\end{figure}

Figures \ref{fi:msenormal} and \ref{fi:deltaformse} display the results. 
The trend of the rMSE in the normalized power prior is generally 
consistent with the findings in a Bernoulli population. 
For the joint power prior approaches, 
the one with the likelihood based on the original data 
is similar to no borrowing. 
The one based on the product of densities using sufficient statistics tends to 
provide some adaptive borrowing, 
while less information is incorporated than 
using the normalized power prior. 
We conclude that the normalized power prior can also 
provide adaptive borrowing under the normal population.


\section{Applications}\label{sec:applications}

\subsection{Water-Quality Assessment}

\begin{figure}[!ht]
\begin{center}
	\psfig{figure=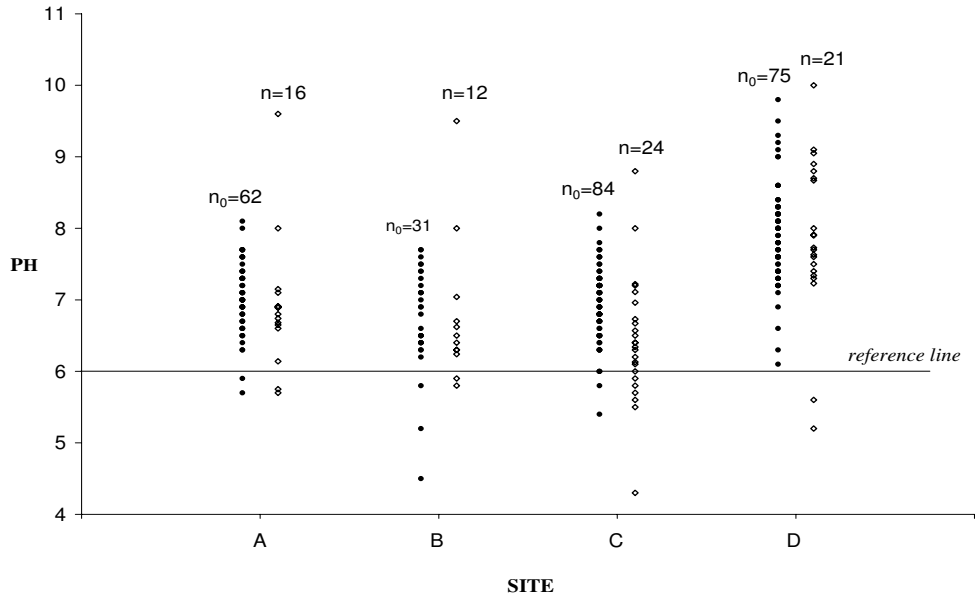,width=5.2 in,height=3.2in} 
\end{center}
\caption{\normalsize
pH data collected at four stations. For each site, historical data
are on the left (circle) and current data on the right (diamond).}
\label{fi:ph}
\end{figure}

In this example, we use
measurements of pH to evaluate impairment of four sites in Virginia
individually. 
pH data collected over a two-year or
three-year period are treated as the current data, while pH data
collected over the previous nine years represents one single
historical data. 
Of interest is the determination of
whether the pH values at a site indicate that the site violates a
(lower) standard of $6.0$ more than $10\%$ of the time. 
For each site, larger sample size is associated with the historical data and
smaller with the current data. 
We apply the normalized power prior
approach, a traditional Bayesian approach for current data only 
using the reference prior, and the joint power prior approaches. 
Assume that the measurements of water quality follow a normal distribution, 
and for ease of comparison, 
the normal model with a simple mean is considered. 
Since the data is used as an illustration to implement the 
normalized power prior, other factors, such as spatial
and temporal features, are not considered.
The current data and historical data are
plotted side by side for each site in Figure \ref{fi:ph}. 
A violation is evaluated using a Bayesian test of
\begin{align*} H_0&:L \geq 6.0~\mbox{(no impairment)},\\ H_1&:L < 6.0~\mbox{(impairment)},
\end{align*}
where $L$ is the lower $10^{th}$ percentile of the distribution for pH.

\begin{table}[!ht]
\caption{
Model fitting results in evaluating 
site impairment with historical data available. In the table 
$n$ and $n_0$ are sample sizes, mean (s.d.) refers to 
sample mean (sample standard
deviation), and s.d. of $L$ is the posterior standard deviation of $L$.} 
\label{PH-Table}
\begin{center}
\begin{adjustbox}{max width=1\textwidth}
\begin{tabular}{l |c c |c c|c |c| c c c}\toprule
 Site& \multicolumn{2}{c|}{Current} & \multicolumn{2}{c|}{Historical} &
  \multicolumn{5}{c}{Posterior probability of $H_0$}\\
 & \multicolumn{2}{c|}{data} & \multicolumn{2}{c|}{data}
 & \multicolumn{5}{c}{(s.d. of $L$)}\\ \cline{2-3}\cline{4-5}\cline{6-10}
 &$n$ &mean  & $n_0$ &mean  &Reference &Normalized  &\multicolumn{3}{c}{Joint power prior}\\
 &$ $ &(s.d.)&       &(s.d.)&  prior   &power prior &(1) &(2) &(3)   \\ \hline
 A &16&6.91  & 62&7.05  &0.177 &0.488 &0.385 & 0.201 &0.997 \\
   &  &(0.90) &&(0.47) &(0.34) &(0.26) &(0.31) & (0.32)& (0.09) \\\hline
 B &12&6.78  &31&6.73  &0.069  &0.047 &0.051 & 0.070 &0.033 \\
   &  &(1.03)&  &(0.71)&(0.47) &(0.26)  &(0.30) & (0.45)& (0.17) \\\hline
 C &24&6.43  &84&6.95  &0.001 &0.004  &0.003 & 0.002 & 0.592\\
   &  &(0.88) & &(0.49)&(0.26) &(0.24)  &(0.25) & (0.25) & (0.08)\\\hline
 D &21&7.87  &75&7.88  &0.865 &0.986  &0.959 & 0.886 & 1.000\\
   &  &(1.11)&  &(0.67)&(0.36) &(0.25)  &(0.30) & (0.35)& (0.11)\\
\bottomrule
\end{tabular}
\end{adjustbox}
\end{center}
\end{table}


Table \ref{PH-Table} summarizes the current and the 
historical data, and the test results using the reference
prior analysis (without incorporating historical data), the 
normalized power prior, and the joint power prior analyses 
(with reference prior as the initial prior for $(\mu,\sigma^2)$, 
i.e, $a=1$ in Section \ref{sec:LM}).  
Similar to Sections \ref{sec:BehaviorComp} and \ref{sec:mse}, 
results from the joint power priors are
calculated using different likelihood functions:
(1) joint density of sufficient statistics; 
(2) product of $n_0$ independent normal densities; 
(3) product of $n_0$ independent normal densities  
multiply by an arbitrary large constant $({2\pi})^{n_0/2}\exp(200)$. 

\begin{figure}[ht!]
\begin{center}
	\psfig{figure=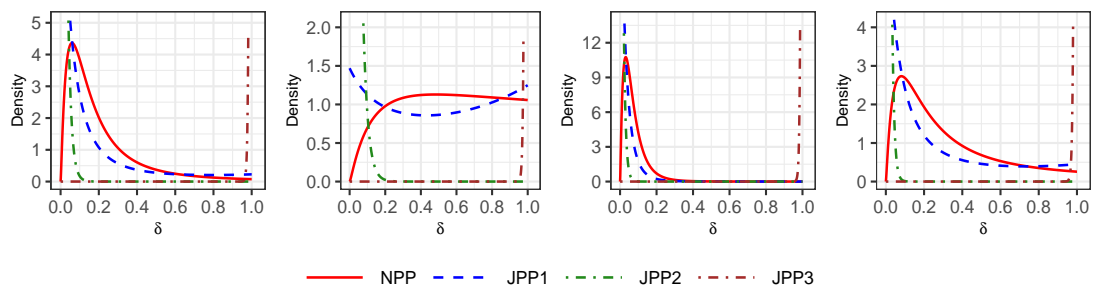,width= 5.7 in,height= 1.6 in} 
\end{center}
\caption{Marginal posterior density plot for $\delta$ using different priors. 
JPP 1 to 3 refer to the joint power priors with different likelihood forms 
as described in the example.}
\label{fi:PH-delta}
\end{figure}

The posterior probability of $H_0$ is calculated based on the posterior 
of $L = \mu + \Phi^{-1}(0.1) \sigma$, where $\Phi^{-1}(\cdot)$ is the 
quantile function of a standard normal distribution. 
If the $0.05$ significance level is used, the Bayesian
test using the reference prior and the 
current data would only indicate site C as
impaired. Here we use the posterior probability of $H_0$ as
equivalent to the p-value \citep{Berger85}. 
Using historical data does lead to different
conclusions for site B. 
The test using normalized power prior results in significance for 
both sites B \& C. 
The test using joint power prior with likelihood (1) 
results in significance for site C, and 
the posterior probability of $H_0$ for site B is very close to $0.05$. 
In the case of site B,
there are around $10\%$ of historical observations below $6.0$.
Hence our prior opinion of the site is suggestive of impairment.
Less information is therefore required to declare impairment
relative to a reference prior and the result is a smaller p-value.
However, if one uses the likelihood function in case (2) of the
joint power prior method, the test result is similar to no borrowing.
Furthermore, if we use an arbitrary constant as in case (3) of the
joint power prior, results will be completely different. 
The standard deviations of $L$ will become very small, 
and it is similar to a full borrowing; see Figure \ref{fi:PH-delta}. 
We will conclude site B impaired, but site C not, 
due to the strong influence of the historical data. 

Hence, this example shows that the inference results are 
sensitive to the likelihood form in 
employing the joint power prior. On the other hand, 
normalized power prior provides adaptive borrowing in all scenarios. 
It is more reasonable to conclude that both site B and site C are impaired.

\subsection{Noninferiority Trial in Vaccine Development }
In a vaccine clinical trial, 
it is commonly required to demonstrate that the 
new vaccine does not interfere with other routine recommended 
vaccines concomitantly.  
In addition to the phase 3 efficacy and safety trials, 
a noninferiority trial is commonly designed 
to demonstrate that the effect (in this example, the response rate) 
of a routine recommended vaccine (vaccine A) 
can be preserved when concomitantly used with the experimental vaccine (vaccine B). 
If the differences in  
the response rate of vaccine A when concomitantly used with 
vaccine B and the response rate of using vaccine A alone 
is within a certain prespecified margin, 
then we may conclude that they do not interfere each other. 
The prespecified positive margin $d_{m}$, known as the noninferiority margin, 
reflects the maximum acceptable extent of clinical noninferiority 
in an experimental treatment. 

A simple frequentist approach of conducting such noninferiority test is to 
calculate the $95\%$ confidence interval of $p_t - p_c$, where 
$p_t$ and $p_c$ are the 
response rates for test and control groups respectively. 
Given a positive noninferiority margin $d_{m}$, 
we conclude that the experimental treatment is not inferior to the control 
if the lower bound of the $95\%$ confidence interval is 
greater than $-d_{m}$. 
When a Bayesian approach is applied, the $95\%$ confidence interval  
can be replaced by the $95\%$ credible interval (CI) based on the 
highest posterior density \citep{Gamalo11}.

However, a problem with either the frequentist or the Bayesian approach 
using noninformative priors is, when the sample size is too small, 
the confidence interval or the credible interval 
will become too wide. 
Therefore inferiority could be inappropriately concluded. 
For this reason, historical evidence, especially historical data 
for the control group, can be incorporated. 
Examples of Bayesian noninferiority trials design based on power prior 
can be found in \cite{Lin16} and \cite{Li18}. 

We illustrate the use of normalized power prior approach to adaptively 
borrow data from historical controls in the development of 
RotaTeq, a live pentavalent rotavirus vaccine. 
A study was designed to investigate the concomitant use of RotaTeq 
and some routine pediatric vaccines between 2001-2005 \citep{Liu18}. 
Specifically, the test was conducted to evaluate the 
anti-polyribosylribitol phosphate response 
(a measure of vaccination against invasive disease caused by 
 Haemophilus influenzae type b) 
to COMVAX (a combination vaccine for Haemophilus influenzae type b 
and hepatitis B), in concomitant use with RotaTeq. 
Since our goal is to assess whether the experimental vaccine RotaTeq 
will affect the response rate of the routine recommended COMVAX or not,
the endpoint is the response rate of COMVAX. 
The per-protocol population included $558$ subjects from the test 
group (COMVAX+RotaTeq) and 592 from the control group 
(COMVAX+placebo). 

Since COMVAX was used for a few years, data from historical trials with similar features 
can be incorporated. 
Table \ref{ta:vaccine} provides a summary of the available datasets 
\citep{Liu18}. 
We pool the four historical data sets, and applying (1) non-informative Bayesian analysis with Jeffrey's prior; 
 (2) joint power prior with the likelihood 
 written as the product of Bernoulli densities, denoted as JPP1; 
(3) joint power prior with likelihood written as the binomial density,  
denoted as JPP2; 
(4) normalized power prior. 
Results are summarized in Table \ref{ta:vaccine2}.

\begin{table}[!ht]
	\begin{center}
		\caption{\em Summary of historical and current studies.}
		\label{ta:vaccine}~ 
		\begin{adjustbox}{max width=1\textwidth}
		\begin{tabular}{ l   c   c   c   c    c  } 
		\toprule
			\multicolumn{2}{l}{Study} & Study Years & N & Responders 
			&Response Rate \\ \hline
			Historical Studies & Study 1 & 1992-1993 & 576 & 417 & 72.4\%  \\ 
			 & Study 2 & 1993-1995 & 111 & 90 & 81.1\%  \\ 
			 & Study 3 & 1993-1995 & 62 & 49 & 79.9\% \\ 
			 & Study 4 & 1997-2000 & 487 & 376 & 77.2\% \\ \midrule
			Current Study & Control & 2001-2005 & 592 & 426 & 72.0\% \\ 
			 & Test & 2001-2005 & 558 & 415 & 74.4\% \\ 
			 \bottomrule
		\end{tabular}
		\end{adjustbox}
	\end{center}
\end{table}

Since the normalized power prior incorporates the most information from the control group 
of the historical studies, 
its 95\% CI of $p_t-p_c$ is the shortest. 
On the other hand, 
using the joint power prior with the product of Bernoulli densities as the likelihood 
results in almost no borrowing, while 
using a binomial density as the likelihood will slightly 
improves the borrowing. 
Since the average response rate in historical controls are slightly larger than that of the   
current control, the estimated response rate of the control group is the 
largest under the normalized power prior. 
This will result in a more conservative decision making when concluding 
noninferiority. 
Under a commonly used noninferiority margin $d_{m} = 5\%$, we can  
conclude noninferiority under all approaches, but in very rare cases, when a smaller margin is chosen, say $d_{m} = 3\%$, 
the noninferiority might be questionable 
when considering more historical information with a  
normalized power prior. 

 The posterior distribution of $\delta$ is skewed, therefore the 
posterior mean is not close to the posterior mode of $\delta$. 
In the normalized power prior approach, 
the posterior mean of $\delta$ is $0.482$, indicating that on average, approximately $1236 \times 48.2\%$ subjects are borrowed from the 
historical data. On the other hand, 
if one considers the power prior with a fixed $\delta$ for ease of interpretation, 
the posterior mode and posterior mean of $\delta$ can serve as the guided values, since they 
provide some useful information regarding the data compatibility. 
For example, considering a fixed $\delta = 0.95$ in 
practice might be anti-conservative, while a fixed $\delta = 0.05$ might be too conservative from the prior-data conflict point of view.

\begin{table}[!ht]
	\begin{center}
		\caption{\em Summary of study results.}
		\label{ta:vaccine2}
		\begin{tabular}{ l   c   c   c   c   }
		\toprule
			Prior & $\hat{p}_{c}~(\%)$  & $95\%$ CI for $p_t-p_c ~(\%)$ 
			& $\bar{\delta}$ & 
			Mode of $\delta$ \\  \midrule
			Jeffrey's Prior & 71.92  &  $(-2.61, 7.58)$ & - & -  \\ 
			JPP1 & 71.93  & $(-2.89, 7.31)$ & 0.001 & 0  \\ 
			JPP2 & 72.68 & $(-3.26, 6.59)$ & 0.166 &  0 \\ 
			NPP & 73.50  & $(-3.76, 5.54)$ &  0.482 &  0.181 \\ 
			\bottomrule
		\end{tabular}
	\end{center}
\end{table}

\subsection{Diagnostic Test Evaluation}
The U.S. Food and Drug Administration (FDA) has released a 
guidance\footnote{the complete version of the guidance 
can be freely downloaded at: 
https://www.fda.gov/media/71512/download [Accessed 03 June 2019].} 
for the use of Bayesian methods in medical device clinical trials.
This guidance specifies that the power prior could be 
one of the  methodologies to borrow strength from other studies. 
In this example, the proposed normalized power prior is 
applied to evaluate the diagnostic test 
for spontaneous preterm delivery (SPD). 
The  binary diagnostic test may result in one of the four possible outcomes: 
true positive (Cell 1), false positive (Cell 2), 
false negative (Cell 3) and true negative (Cell 4); 
see Table \ref{pc}. 
Let $\boldsymbol{\theta}=(\theta_1,\theta_2,\theta_3,\theta_4)$ denote the cell probabilities and let $\boldsymbol{n}=(n_1,n_2,n_3,n_4)$ denote the corresponding number of subjects in Table \ref{pc}. 
The {\it sensitivity} $\eta$ and {\it specificity}
$\lambda$ of a test can be expressed in terms of the cell probabilities $\boldsymbol{\theta}$ as
\[
\eta\equiv \textbf{Pr}(T^+\mid D^+)\equiv\frac{\theta_1}{\theta_1+\theta_3},\quad\text{and}\quad \lambda\equiv\textbf{Pr}(T^-\mid D^-)\equiv\frac{\theta_4}{\theta_2+\theta_4},
\]
respectively, where $D$ stands for disease status and 
$T$ stands for test status. 

\begin{table}[ht!]
\begin{center}
\caption{\em Possible outcomes of a binary diagnostic test.}
\begin{tabular}{lclcl}
\toprule
 &  & \multicolumn{3}{c}{Disease status}     \\ \cline{3-5} 
                      &  & \multicolumn{1}{c}{Yes}              &  & \multicolumn{1}{c}{No}                \\ \hline
Test positive         &  & Cell 1
 (TP)  &  & Cell 2
 (FP) \\
Test negative         &  & Cell 3 
 (FN) &  & Cell 4 
 (TN) \\ \bottomrule
\end{tabular}
\label{pc}
\end{center}
\end{table}

A simple frequentist approach to evaluate such binary test is to 
compute the 95\% confidence intervals of $\eta$ and $\lambda$, 
denoted by $(\eta_L,\eta_U)$ and $(\lambda_L,\lambda_U)$. 
Then we compare the lower bounds $\eta_L$ and $\lambda_L$ to the 
value of 50\% which is the sensitivity and specificity of a random test.
We may conclude that the diagnostic test outweighs a  
random test on the diseased group if $\eta_L$ is greater than 50\%. 
Similarly, the diagnostic test outweighs a random test 
on non-diseased group if $\lambda_L$ is greater than 50\%.

In practice, however, the diseased group's data are difficult to collect leading to a relatively small $n_1+n_3$. 
As a result, the confidence interval of $\eta$ tends to be 
too wide to make any conclusions. 
For the purpose of this agreement, the sequential Bayesian updating and the power prior can be used to incorporate the historical/external information. 

 A diagnostic test based on a medical device (PartoSure Test-P160052) 
 was developed to aid in rapidly assess the risk of spontaneous preterm 
 delivery within 7 days from the time of diagnosis in pre-pregnant women with signs and symptoms\footnote{the dataset used in this example is freely available at:
 https://www.accessdata.fda.gov/cdrh\_docs/pdf16/P160052C.pdf [Accessed 03 June 2019].}. 
 Table \ref{example_data} lists the dataset of 686 subjects from  the US study and the dataset of 511 subjects from the European study.  
 The test was approved by FDA based on the US study, so  
 the European study is regarded as the external information 
 in this example. 
 The joint power prior (with the full multinomial likelihood), 
 the normalized power prior, 
 no borrowing and full borrowing are applied, with Jeffrey's prior $\text{Dir}(0.5,0.5,0.5,0.5)$ as the initial prior for $\boldsymbol{\theta}$. 
 Table \ref{example_result} summarizes the results.  
 It is found that the posterior mean under the power prior is always between the posterior mean of no borrowing and full borrowing. 
 Also, the result of using joint power prior is close to the one of no 
 borrowing since only $4.4\%$ of the external information is incorporated on average. 
 Using the normalized power prior will on average increase the involved external information to $21.6\%$, 
 making its result closer to  the full borrowing. 
 In practice, the posterior mean of $\delta$ 
 (e.g, $4.4\%$ and $21.6\%$) could be important to clinicians because it not only reflects the information amount that is borrowed, 
 but also indicates the average sample size (e.g., $511\times 4.4\%$ and $511\times 21.6\%$) that is incorporated. 
 The joint power prior suggests very little borrowing while the normalized power prior suggests a moderate level of borrowing. 
 In general, 
 these two data sets are compatible since they have similar sensitivity ($50\%$ and $50\%$)  and specificity ($96\%$ and $98\%$).
 The value obtained by the normalized power prior 
 is more persuasive and reflects the data compatibility.

\begin{table}[!ht]
\begin{center}
\caption{\em 2$\times$2 performance tables with the US study and the European study.}
\begin{adjustbox}{max width=1\textwidth}
\begin{tabular}{llccccclllccccc}
\toprule
US study      &  & \multicolumn{3}{c}{Disease status} &  &       &  & European study &  & \multicolumn{3}{c}{Disease status} &  &       \\ \cline{3-5} \cline{11-13}
              &  & Yes         &         & No         &  & Total &  &               &  & Yes         &         & No         &  & Total \\ \hline
Test positive &  & 3          &         & 11        &  & 14   &  & Test positive &  & 9         &         & 20         &  & 29   \\
Test negative &  & 3           &         & 669        &  & 672   &  & Test negative &  & 9         &         & 473       &  & 482  \\
Total         &  & 6          &         & 680       &  & 686  &  & Total         &  & 18          &         & 493        &  & 511  \\ \bottomrule
\end{tabular}
\end{adjustbox}
\label{example_data}
\end{center}
\end{table}

\begin{table}[!ht]
	\begin{center}
     \caption{\em Summary of study results.}
     \begin{adjustbox}{max width=1\textwidth}
		\begin{tabular}{lcccccc}
\toprule
Prior & $100\hat{\eta}$ & 95\% CI for $\eta$ (\%) & 
$100\hat{\lambda}$ & 95\% CI for $\lambda$ (\%)& 
$\bar{\delta}$  & Mode of  $\delta$ \\ \midrule
Fixed $\delta=0$  &      50.04        &   (16.67, 82.80)                 &              98.31  &      (97.32, 99.22)               &         -                                     &   -                             \\
Fixed $\delta=1$   &   49.85           &      (31.40, 68.70)              &  97.32              &      (96.38, 98.17)                &                                            -  &   -                             \\
JPP                &    49.98        &    (18.94, 83.05)                &              98.24  &      (97.27, 99.18)                &         0.044                                     &  0                              \\
NPP                &      49.88        &         (21.60, 78.84)           &  98.02              &     (96.93, 99.00)                &     0.216                                         &                0.085                \\ \bottomrule
\end{tabular}
\end{adjustbox}
\label{example_result}
	\end{center}
\end{table}

\section{Summary and Discussion}\label{sec:discussion}

As a general class of the informative priors for Bayesian
inference, the power prior provides a framework to incorporate data from
alternative sources, whose influence on statistical inference can be  
adjusted according to its availability and its discrepancy between the current data. 
It is semi-automatic, in the sense that it takes the
form of raising the likelihood function based on the historical data
to a fractional power regardless of the specific form of
heterogeneity. 
As a consequence of using more data, the power prior has 
advantages in terms of the estimation    
with small sample sizes. 
When we do not have enough knowledge 
to model such heterogeneity and cannot specify a fixed power parameter 
in advance, a power prior with a random $\delta$ is 
especially attractive in practice. 

In this article we provide a framework of using 
the normalized power prior approach, 
in which the degree of borrowing is dynamically 
adjusted through the prior-data conflict. 
The subjective information about the difference in two populations 
can be incorporated 
by adjusting the hyperparameters in the prior for $\delta$, and the
discrepancy between the two samples is automatically taken into account
through a random $\delta$.
Theoretical justification is provided based on the 
weighted KL divergence. 
The controlling role of the power parameter in the
normalized power prior is adjusted automatically based on the 
congruence between the historical and the current samples and their sample sizes; 
this is shown using both the analytical and numerical results. 
On the other hand, we revisit some undesirable properties of using 
the joint power prior for a random $\delta$; 
this is shown by theoretical justifications and graphical examples.   
Efficient algorithms for posterior sampling using the 
normalized power prior are also discussed and implemented. 

We acknowledge when $\delta$ is considered random and estimated with a 
Bayesian approach, the normalized power prior is more appropriate. 
The violation of likelihood principle under the 
joint power prior was discussed in 
\cite{Duan06a} and \cite{Neuenschwander09}.
However, a comprehensive study on the joint power prior and 
the normalized power prior is not available in literature. 
As a result, the joint power priors with random $\delta$ 
were still used afterwards, for example, 
\cite{Zhao14}, \cite{Gamalo14}, \cite{Lin16}, and \cite{Zhang19}.
This might partially due to the fact that 
the undesirable behavior of the joint power priors 
were not fully studied and recognized. 
Although under certain 
likelihood forms, the joint power priors  
would provide limited adaptive borrowing, 
its mechanism is unclear. 
We conclude that the joint power prior is not recommended with a random $\delta$.


On the other hand, the power prior with $\delta$ fixed 
is widely used in both clinical trial design and observational 
studies. 
It can be viewed as a 
special case of the normalized power prior with initial prior of 
$\delta$ coming from a degenerate distribution. 
We conjecture that a similar sensitivity analysis used in a 
power prior with  $\delta$ fixed \citep{Ibrahim15}
might be carried out to search for the initial prior of $\delta$ 
in the normalized power prior context. 
Since the normalized power prior generalizes the power prior with 
$\delta$ fixed, 
most inferential results in power prior with $\delta$ fixed 
could be easily adopted. Further studies will be carried out elsewhere.

\section*{Disclaimer}
This article represents the views of the authors and should not be construed to represent FDA's views or policies.

\section*{Acknowledgements}
We warmly thank the anonymous referees and the associate editor for 
helpful comments and suggestions that lead to an improved article.
This work is partially supported by ``the Fundamental 
Research Funds for the Central Universities" in UIBE(CXTD11-05) and a research grant by College of Business at University of Texas at San Antonio. 

\bibliographystyle{elsarticle-num-names}
\biboptions{authoryear}
\bibliography{nppbib.bib}

\begin{thebibliography}{36}
\expandafter\ifx\csname natexlab\endcsname\relax\def\natexlab#1{#1}\fi
\providecommand{\url}[1]{\texttt{#1}}
\providecommand{\href}[2]{#2}
\providecommand{\path}[1]{#1}
\providecommand{\DOIprefix}{doi:}
\providecommand{\ArXivprefix}{arXiv:}
\providecommand{\URLprefix}{URL: }
\providecommand{\Pubmedprefix}{pmid:}
\providecommand{\doi}[1]{\href{http://dx.doi.org/#1}{\path{#1}}}
\providecommand{\Pubmed}[1]{\href{pmid:#1}{\path{#1}}}
\providecommand{\bibinfo}[2]{#2}
\ifx\xfnm\relax \def\xfnm[#1]{\unskip,\space#1}\fi
\bibitem[{Ibrahim and Chen(1998)}]{IbrahimChen98}
\bibinfo{author}{J.~G. Ibrahim}, \bibinfo{author}{M.-H. Chen},
\newblock \bibinfo{title}{Prior distributions and bayesian computation for
  proportional hazards models},
\newblock \bibinfo{journal}{Sankhya: The Indian Journal of Statistics, Series
  B}  (\bibinfo{year}{1998}) \bibinfo{pages}{48--64}.
\bibitem[{Chen et~al.(2000)Chen, Ibrahim, and Shao}]{Chen00}
\bibinfo{author}{M.-H. Chen}, \bibinfo{author}{J.~G. Ibrahim},
  \bibinfo{author}{Q.-M. Shao},
\newblock \bibinfo{title}{Power prior distributions for generalized linear
  models},
\newblock \bibinfo{journal}{Journal of Statistical Planning and Inference}
  \bibinfo{volume}{84} (\bibinfo{year}{2000}) \bibinfo{pages}{121--137}.
\bibitem[{Ibrahim and Chen(2000)}]{IbrahimChen00}
\bibinfo{author}{J.~G. Ibrahim}, \bibinfo{author}{M.-H. Chen},
\newblock \bibinfo{title}{Power prior distributions for regression models},
\newblock \bibinfo{journal}{Statistical Science} \bibinfo{volume}{15}
  (\bibinfo{year}{2000}) \bibinfo{pages}{46--60}.
\bibitem[{Ibrahim et~al.(2003)Ibrahim, Chen, and Sinha}]{Ibrahim03}
\bibinfo{author}{J.~G. Ibrahim}, \bibinfo{author}{M.-H. Chen},
  \bibinfo{author}{D.~Sinha},
\newblock \bibinfo{title}{On optimality properties of the power prior},
\newblock \bibinfo{journal}{Journal of the American Statistical Association}
  \bibinfo{volume}{98} (\bibinfo{year}{2003}) \bibinfo{pages}{204--213}.
\bibitem[{Chen and Ibrahim(2006)}]{Chen06}
\bibinfo{author}{M.-H. Chen}, \bibinfo{author}{J.~G. Ibrahim},
\newblock \bibinfo{title}{The relationship between the power prior and
  hierarchical models},
\newblock \bibinfo{journal}{Bayesian Analysis} \bibinfo{volume}{1}
  (\bibinfo{year}{2006}) \bibinfo{pages}{551--574}.
\bibitem[{Ibrahim et~al.(2015)Ibrahim, Chen, Gwon, and Chen}]{Ibrahim15}
\bibinfo{author}{J.~G. Ibrahim}, \bibinfo{author}{M.-H. Chen},
  \bibinfo{author}{Y.~Gwon}, \bibinfo{author}{F.~Chen},
\newblock \bibinfo{title}{The power prior: Theory and applications},
\newblock \bibinfo{journal}{Statistics in Medicine} \bibinfo{volume}{34}
  (\bibinfo{year}{2015}) \bibinfo{pages}{3724--3749}.
\bibitem[{Ibrahim et~al.(2012)Ibrahim, Chen, and Chu}]{Ibrahim12}
\bibinfo{author}{J.~G. Ibrahim}, \bibinfo{author}{M.-H. Chen},
  \bibinfo{author}{H.~Chu},
\newblock \bibinfo{title}{Bayesian methods in clinical trials: a bayesian
  analysis of ecog trials e1684 and e1690},
\newblock \bibinfo{journal}{BMC Medical Research Methodology}
  \bibinfo{volume}{12} (\bibinfo{year}{2012}) \bibinfo{pages}{183}.
\bibitem[{Spiegelhalter et~al.(2002)Spiegelhalter, Best, Carlin, and Van
  Der~Linde}]{Spiegelhalter02}
\bibinfo{author}{D.~J. Spiegelhalter}, \bibinfo{author}{N.~G. Best},
  \bibinfo{author}{B.~P. Carlin}, \bibinfo{author}{A.~Van Der~Linde},
\newblock \bibinfo{title}{Bayesian measures of model complexity and fit},
\newblock \bibinfo{journal}{Journal of the Royal Statistical Society: Series B}
  \bibinfo{volume}{64} (\bibinfo{year}{2002}) \bibinfo{pages}{583--639}.
\bibitem[{Birnbaum(1962)}]{Birnbaum62}
\bibinfo{author}{A.~Birnbaum},
\newblock \bibinfo{title}{On the foundations of statistical inference},
\newblock \bibinfo{journal}{Journal of the American Statistical Association}
  \bibinfo{volume}{57} (\bibinfo{year}{1962}) \bibinfo{pages}{269--306}.
\bibitem[{Neuenschwander et~al.(2009)Neuenschwander, Branson, and
  Spiegelhalter}]{Neuenschwander09}
\bibinfo{author}{B.~Neuenschwander}, \bibinfo{author}{M.~Branson},
  \bibinfo{author}{D.~J. Spiegelhalter},
\newblock \bibinfo{title}{A note on the power prior},
\newblock \bibinfo{journal}{Statistics in Medicine} \bibinfo{volume}{28}
  (\bibinfo{year}{2009}) \bibinfo{pages}{3562--3566}.
\bibitem[{Neelon and O'Malley(2010)}]{Neelon10}
\bibinfo{author}{B.~Neelon}, \bibinfo{author}{A.~O'Malley},
\newblock \bibinfo{title}{Bayesian analysis using power priors with application
  to pediatric quality of care},
\newblock \bibinfo{journal}{Journal of Biometrics \& Biostatistics}
  \bibinfo{volume}{1} (\bibinfo{year}{2010}) \bibinfo{pages}{103}.
\bibitem[{Duan et~al.(2006)Duan, Ye, and Smith}]{Duan06a}
\bibinfo{author}{Y.~Duan}, \bibinfo{author}{K.~Ye}, \bibinfo{author}{E.~P.
  Smith},
\newblock \bibinfo{title}{Evaluating water quality using power priors to
  incorporate historical information},
\newblock \bibinfo{journal}{Environmetrics} \bibinfo{volume}{17}
  (\bibinfo{year}{2006}) \bibinfo{pages}{95--106}.
\bibitem[{Gamalo et~al.(2014)Gamalo, Tiwari, and LaVange}]{Gamalo14}
\bibinfo{author}{M.~A. Gamalo}, \bibinfo{author}{R.~C. Tiwari},
  \bibinfo{author}{L.~M. LaVange},
\newblock \bibinfo{title}{Bayesian approach to the design and analysis of
  non-inferiority trials for anti-infective products},
\newblock \bibinfo{journal}{Pharmaceutical Statistics} \bibinfo{volume}{13}
  (\bibinfo{year}{2014}) \bibinfo{pages}{25--40}.
\bibitem[{Gravestock and Held(2019)}]{Gravestock18}
\bibinfo{author}{I.~Gravestock}, \bibinfo{author}{L.~Held},
\newblock \bibinfo{title}{Power priors based on multiple historical studies for
  binary outcomes},
\newblock \bibinfo{journal}{Biometrical Journal}  (\bibinfo{year}{2019})
  \bibinfo{pages}{1201--1218}.
\bibitem[{Banbeta et~al.(2019)Banbeta, van Rosmalen, Dejardin, and
  Lesaffre}]{Banbeta19}
\bibinfo{author}{A.~Banbeta}, \bibinfo{author}{J.~van Rosmalen},
  \bibinfo{author}{D.~Dejardin}, \bibinfo{author}{E.~Lesaffre},
\newblock \bibinfo{title}{Modified power prior with multiple historical trials
  for binary endpoints},
\newblock \bibinfo{journal}{Statistics in Medicine} \bibinfo{volume}{38}
  (\bibinfo{year}{2019}) \bibinfo{pages}{1147--1169}.
\bibitem[{Ibrahim et~al.(2012)Ibrahim, Chen, Xia, and Liu}]{Ibrahim12-2}
\bibinfo{author}{J.~G. Ibrahim}, \bibinfo{author}{M.-H. Chen},
  \bibinfo{author}{H.~A. Xia}, \bibinfo{author}{T.~Liu},
\newblock \bibinfo{title}{Bayesian meta-experimental design: Evaluating
  cardiovascular risk in new antidiabetic therapies to treat type 2 diabetes},
\newblock \bibinfo{journal}{Biometrics} \bibinfo{volume}{68}
  (\bibinfo{year}{2012}) \bibinfo{pages}{578--586}.
\bibitem[{Chen et~al.(2014{\natexlab{a}})Chen, Ibrahim, Zeng, Hu, and
  Jia}]{Chen14-2}
\bibinfo{author}{M.-H. Chen}, \bibinfo{author}{J.~G. Ibrahim},
  \bibinfo{author}{D.~Zeng}, \bibinfo{author}{K.~Hu}, \bibinfo{author}{C.~Jia},
\newblock \bibinfo{title}{Bayesian design of superiority clinical trials for
  recurrent events data with applications to bleeding and transfusion events in
  myelodyplastic syndrome},
\newblock \bibinfo{journal}{Biometrics} \bibinfo{volume}{70}
  (\bibinfo{year}{2014}{\natexlab{a}}) \bibinfo{pages}{1003--1013}.
\bibitem[{Chen et~al.(2014{\natexlab{b}})Chen, Ibrahim, Xia, Liu, and
  Hennessey}]{Chen14}
\bibinfo{author}{M.-H. Chen}, \bibinfo{author}{J.~G. Ibrahim},
  \bibinfo{author}{H.~A. Xia}, \bibinfo{author}{T.~Liu},
  \bibinfo{author}{V.~Hennessey},
\newblock \bibinfo{title}{Bayesian sequential meta-analysis design in
  evaluating cardiovascular risk in a new antidiabetic drug development
  program},
\newblock \bibinfo{journal}{Statistics in Medicine} \bibinfo{volume}{33}
  (\bibinfo{year}{2014}{\natexlab{b}}) \bibinfo{pages}{1600--1618}.
\bibitem[{Chib and Greenberg(1995)}]{Chib95}
\bibinfo{author}{S.~Chib}, \bibinfo{author}{E.~Greenberg},
\newblock \bibinfo{title}{Understanding the metropolis-hastings algorithm},
\newblock \bibinfo{journal}{The American Statistician} \bibinfo{volume}{49}
  (\bibinfo{year}{1995}) \bibinfo{pages}{327--335}.
\bibitem[{Friel and Pettitt(2008)}]{Friel08}
\bibinfo{author}{N.~Friel}, \bibinfo{author}{A.~N. Pettitt},
\newblock \bibinfo{title}{Marginal likelihood estimation via power posteriors},
\newblock \bibinfo{journal}{Journal of the Royal Statistical Society: Series B
  (Statistical Methodology)} \bibinfo{volume}{70} (\bibinfo{year}{2008})
  \bibinfo{pages}{589--607}.
\bibitem[{Van~Rosmalen et~al.(2018)Van~Rosmalen, Dejardin, van Norden,
  L{\"o}wenberg, and Lesaffre}]{Rosmalen18}
\bibinfo{author}{J.~Van~Rosmalen}, \bibinfo{author}{D.~Dejardin},
  \bibinfo{author}{Y.~van Norden}, \bibinfo{author}{B.~L{\"o}wenberg},
  \bibinfo{author}{E.~Lesaffre},
\newblock \bibinfo{title}{Including historical data in the analysis of clinical
  trials: Is it worth the effort?},
\newblock \bibinfo{journal}{Statistical Methods in Medical Research}
  \bibinfo{volume}{27} (\bibinfo{year}{2018}) \bibinfo{pages}{3167--3182}.
\bibitem[{Gelman and Meng(1998)}]{GelmanMeng98}
\bibinfo{author}{A.~Gelman}, \bibinfo{author}{X.-L. Meng},
\newblock \bibinfo{title}{Simulating normalizing constants: From importance
  sampling to bridge sampling to path sampling},
\newblock \bibinfo{journal}{Statistical Science} \bibinfo{volume}{13}
  (\bibinfo{year}{1998}) \bibinfo{pages}{163--185}.
\bibitem[{Carvalho and Ibrahim(2020)}]{Carvalho20}
\bibinfo{author}{L.~M. Carvalho}, \bibinfo{author}{J.~G. Ibrahim},
\newblock \bibinfo{title}{On the normalized power prior},
\newblock \bibinfo{journal}{arXiv:2004.14912} \bibinfo{volume}{v1}
  (\bibinfo{year}{2020}) \bibinfo{pages}{1--31}.
\bibitem[{Casella and Berger(2002)}]{CasellaBerger01}
\bibinfo{author}{G.~Casella}, \bibinfo{author}{R.~L. Berger},
  \bibinfo{title}{Statistical Inference}, \bibinfo{edition}{2nd} ed.,
  \bibinfo{publisher}{Duxbury Pacific Grove, CA}, \bibinfo{year}{2002}.
\bibitem[{Zellner(1986)}]{Zellner86}
\bibinfo{author}{A.~Zellner}, \bibinfo{title}{On Assessing Prior Distributions
  and Bayesian Regression Analysis with g-Prior Distributions},
  \bibinfo{publisher}{Elsevier, New York.}, \bibinfo{year}{1986}.
\bibitem[{Kullback and Leibler(1951)}]{Kullback51}
\bibinfo{author}{S.~Kullback}, \bibinfo{author}{R.~A. Leibler},
\newblock \bibinfo{title}{On information and sufficiency},
\newblock \bibinfo{journal}{The Annals of Mathematical Statistics}
  \bibinfo{volume}{22} (\bibinfo{year}{1951}) \bibinfo{pages}{79--86}.
\bibitem[{Berger and Bernardo(1992)}]{BergerBernardo92}
\bibinfo{author}{J.~O. Berger}, \bibinfo{author}{J.~M. Bernardo},
  \bibinfo{title}{On the development of reference priors},
  \bibinfo{year}{1992}.
\bibitem[{Berger(2013)}]{Berger85}
\bibinfo{author}{J.~O. Berger}, \bibinfo{title}{Statistical Decision Theory and
  Bayesian Analysis}, \bibinfo{publisher}{Springer Science \& Business Media},
  \bibinfo{year}{2013}.
\bibitem[{Gamalo et~al.(2011)Gamalo, Wu, and Tiwari}]{Gamalo11}
\bibinfo{author}{M.~A. Gamalo}, \bibinfo{author}{R.~Wu}, \bibinfo{author}{R.~C.
  Tiwari},
\newblock \bibinfo{title}{Bayesian approach to noninferiority trials for
  proportions},
\newblock \bibinfo{journal}{Journal of Biopharmaceutical Statistics}
  \bibinfo{volume}{21} (\bibinfo{year}{2011}) \bibinfo{pages}{902--919}.
\bibitem[{Lin et~al.(2016)Lin, Gamalo-Siebers, and Tiwari}]{Lin16}
\bibinfo{author}{J.~Lin}, \bibinfo{author}{M.~Gamalo-Siebers},
  \bibinfo{author}{R.~Tiwari},
\newblock \bibinfo{title}{Non-inferiority and networks: Inferring efficacy from
  a web of data},
\newblock \bibinfo{journal}{Pharmaceutical Statistics} \bibinfo{volume}{15}
  (\bibinfo{year}{2016}) \bibinfo{pages}{54--67}.
\bibitem[{Li et~al.(2018)Li, Chen, Wang, and Dey}]{Li18}
\bibinfo{author}{W.~Li}, \bibinfo{author}{M.-H. Chen},
  \bibinfo{author}{X.~Wang}, \bibinfo{author}{D.~K. Dey},
\newblock \bibinfo{title}{Bayesian design of non-inferiority clinical trials
  via the bayes factor},
\newblock \bibinfo{journal}{Statistics in Biosciences} \bibinfo{volume}{10}
  (\bibinfo{year}{2018}) \bibinfo{pages}{439--459}.
\bibitem[{Liu(2018)}]{Liu18}
\bibinfo{author}{G.~F. Liu},
\newblock \bibinfo{title}{A dynamic power prior for borrowing historical data
  in noninferiority trials with binary endpoint},
\newblock \bibinfo{journal}{Pharmaceutical Statistics} \bibinfo{volume}{17}
  (\bibinfo{year}{2018}) \bibinfo{pages}{61--73}.
\bibitem[{Zhao et~al.(2014)Zhao, Zalkikar, Tiwari, and LaVange}]{Zhao14}
\bibinfo{author}{Y.~Zhao}, \bibinfo{author}{J.~Zalkikar},
  \bibinfo{author}{R.~C. Tiwari}, \bibinfo{author}{L.~M. LaVange},
\newblock \bibinfo{title}{A bayesian approach for benefit-risk assessment},
\newblock \bibinfo{journal}{Statistics in Biopharmaceutical Research}
  \bibinfo{volume}{6} (\bibinfo{year}{2014}) \bibinfo{pages}{326--337}.
\bibitem[{Zhang et~al.(2019)Zhang, Ko, Nie, Chen, and Tiwari}]{Zhang19}
\bibinfo{author}{J.~Zhang}, \bibinfo{author}{C.-W. Ko},
  \bibinfo{author}{L.~Nie}, \bibinfo{author}{Y.~Chen}, \bibinfo{author}{R.~C.
  Tiwari},
\newblock \bibinfo{title}{Bayesian hierarchical methods for meta-analysis
  combining randomized-controlled and single-arm studies},
\newblock \bibinfo{journal}{Statistical Methods in Medical Research}
  \bibinfo{volume}{28} (\bibinfo{year}{2019}) \bibinfo{pages}{1293--1310}.
\bibitem[{Gelman et~al.(2013)Gelman, Carlin, Stern, Dunson, Vehtari, and
  Rubin}]{Gelman13}
\bibinfo{author}{A.~Gelman}, \bibinfo{author}{J.~B. Carlin},
  \bibinfo{author}{H.~S. Stern}, \bibinfo{author}{D.~B. Dunson},
  \bibinfo{author}{A.~Vehtari}, \bibinfo{author}{D.~B. Rubin},
  \bibinfo{title}{Bayesian Data Analysis}, \bibinfo{edition}{3rd} ed.,
  \bibinfo{publisher}{Chapman and Hall/CRC}, \bibinfo{year}{2013}.
\bibitem[{Carpenter et~al.(2017)Carpenter, Gelman, Hoffman, Lee, Goodrich,
  Betancourt, Brubaker, Guo, Li, and Riddell}]{Carpenter17}
\bibinfo{author}{B.~Carpenter}, \bibinfo{author}{A.~Gelman},
  \bibinfo{author}{M.~D. Hoffman}, \bibinfo{author}{D.~Lee},
  \bibinfo{author}{B.~Goodrich}, \bibinfo{author}{M.~Betancourt},
  \bibinfo{author}{M.~Brubaker}, \bibinfo{author}{J.~Guo},
  \bibinfo{author}{P.~Li}, \bibinfo{author}{A.~Riddell},
\newblock \bibinfo{title}{Stan: A probabilistic programming language},
\newblock \bibinfo{journal}{Journal of Statistical Software}
  \bibinfo{volume}{76} (\bibinfo{year}{2017}) \bibinfo{pages}{1--32}.

\end{thebibliography}







\appendix
\section{Proofs and Theorems}\label{appa}

\noindent \textbf{Proof of Identity (\ref{eq:calpha}):}\\
\vspace*{-0.15in}
Taking derivative of  
$\log \int_{\boldsymbol{\Theta}} L(\boldsymbol{\theta}|D_{0})^\delta 
\pi_{0}(\boldsymbol{\theta}) 
d\boldsymbol{\theta}$ with respect to $\delta$ we have: 
\begin{align*}
\frac{d}{d \delta} \log \int_{\boldsymbol{\Theta}} &L(\boldsymbol{\theta}|D_{0})^\delta 
\pi_{0}(\boldsymbol{\theta}) 
d\boldsymbol{\theta} = 
\frac{1}{\int_{\boldsymbol{\Theta}} L(\boldsymbol{\theta}|D_{0})^\delta 
\pi_{0}(\boldsymbol{\theta}) 
d\boldsymbol{\theta}} 
\frac{d}{d \delta} \int_{\boldsymbol{\Theta}} L(\boldsymbol{\theta}|D_{0})^\delta 
\pi_{0}(\boldsymbol{\theta}) 
d\boldsymbol{\theta} \\
&= \frac{1}{\int_{\boldsymbol{\Theta}} L(\boldsymbol{\theta}|D_{0})^\delta 
\pi_{0}(\boldsymbol{\theta}) 
d\boldsymbol{\theta}}
\int_{\boldsymbol{\Theta}} L(\boldsymbol{\theta}|D_{0})^\delta 
\log [L(\boldsymbol{\theta}|D_{0})]
\pi_{0}(\boldsymbol{\theta}) 
d\boldsymbol{\theta}\\
&= 
\int_{\boldsymbol{\Theta}} 
\frac{
L(\boldsymbol{\theta}|D_{0})^\delta 
\pi_{0}(\boldsymbol{\theta})}{
\int_{\boldsymbol{\Theta}} L(\boldsymbol{\theta}|D_{0})^\delta 
\pi_{0}(\boldsymbol{\theta}) d\boldsymbol{\theta}
}
\log [L(\boldsymbol{\theta}|D_{0})] 
d\boldsymbol{\theta} \\
&= 
E_{\pi(\boldsymbol{\theta} | D_0, \delta )} 
\{ \log [ L( \boldsymbol{\theta}|D_0 ) ] \}.
\end{align*}
So the equation (\ref{eq:calpha}) can be obtained by 
integrating with respect to $\delta$. \\

\noindent 
\textbf{Proof of Theorem \ref{thm:LM}:}
To prove the Theorem \ref{thm:LM}, we first state two simple identities of linear algebra and multivariate integral without proof. 
For positive-definite $k\times k$ matrices 
$\boldsymbol{A}$ and 
$\boldsymbol{B}$, and $k\times 1$ vectors $\boldsymbol{x}$, $\boldsymbol{y}$, and $\boldsymbol{z}$,
\begin{align}\label{eq:identity1}
    &(\boldsymbol{x}-\boldsymbol{y})'
    \boldsymbol{A}(\boldsymbol{x}-\boldsymbol{y}) + (\boldsymbol{x}-\boldsymbol{z})'
    \boldsymbol{B}(\boldsymbol{x}-\boldsymbol{z}) = (\boldsymbol{y}-\boldsymbol{z})'\boldsymbol{B}
    (\boldsymbol{A+B})^{-1}\boldsymbol{A} 
    (\boldsymbol{y}-\boldsymbol{z})\nonumber\\ &+ \left[\boldsymbol{x}-(
    \boldsymbol{A+B})^{-1}(\boldsymbol{Ay+Bz})\right]'
    (\boldsymbol{A+B})
    \left[\boldsymbol{x}-
    (\boldsymbol{A+B})^{-1}(\boldsymbol{Ay+Bz})\right].
\end{align}
On the other hand, for $\boldsymbol{A}$ being a positive-definite 
$k\times k$ matrix, 
$\boldsymbol{x}$ and $\boldsymbol{x}_0$ $k\times 1$ vectors, with positive constants $t$, $a$ and $b$ where $a > \frac{k}{2}+1$,
\begin{align}\label{eq:identity2}
&\int_0^{\infty}\int_{\mathcal{R}^k} \frac{1}{t^a}
\exp \left\{ -\frac{b+(\boldsymbol{x}-\boldsymbol{x}_0)'
\boldsymbol{A}(\boldsymbol{x}-\boldsymbol{x}_0)}{2t} \right\}
d\boldsymbol{x}dt \nonumber\\ &= \left(2\pi \right)^{\frac{k}{2}}\Gamma\left(a-\frac{k}{2}-1\right)
|\boldsymbol{A}|^{-\frac{1}{2}} \left(\frac{b}{2}\right)^
{-\left(a-\frac{k}{2}-1\right)}.
\end{align}

For the current data $D$, the likelihood function of $(\boldsymbol{\beta},\sigma^2)$ using (\ref{eq:LM}) can be written as 
$$L(\boldsymbol{\beta},\sigma^2|D)\propto 
 \frac{1}{(\sigma^2)^{\frac{n}{2}}}\exp\left\{
- \frac{1}{2 \sigma^2} \left[
S + (\boldsymbol{\beta}-\hat{\boldsymbol{\beta}})'\mathbf{X}'\mathbf{X}(\boldsymbol{\beta}-\hat{\boldsymbol{\beta}})
\right] \right\},
$$
where ${S}$ is defined in Section \ref{sec:LM}.  
Accordingly, adding subscript $0$ to data and all other quantities except for the 
parameters $(\boldsymbol{\beta},\sigma^2)$ would give similar form to $L(\boldsymbol{\beta},\sigma^2|D_0)$.
\begin{itemize}
    \item[(a)] To obtain the normalized power prior, we need to find the normalization factor
    \begin{align*}
        &C(\delta) \propto \int_{0}^{\infty} \int_{\mathcal{R}^k} \pi_{0}(\boldsymbol{\beta},\sigma^2)L(\boldsymbol{\beta},\sigma^2|D_0)^{\delta}d\boldsymbol{\beta}d\sigma^2 \\
        &\propto \int_{0}^{\infty} \int_{\mathcal{R}^k}  \frac{1}{(\sigma^2)^{\frac{\delta n_0 + bk}{2}+a}}
\exp\left\{-\frac{1}{2\sigma^2}\left[\delta\left\{{S}_0 + b{H}_0(\delta)\right\} +  {Q}(\delta,\boldsymbol{\beta})\right]\right\} d\boldsymbol{\beta}d\sigma^2
\\&\propto 
1/M_0(\delta), 
\end{align*}
where $H_0(\delta)$, $M_0(\delta)$ and $Q(\delta, \boldsymbol{\beta})$ are defined in Theorem \ref{thm:LM} (a). Note that, using (\ref{eq:identity1}),
the second line follows from completing the squares 
$$
(\boldsymbol{\beta} -{\boldsymbol{\mu}}_{0})' b \boldsymbol{R} (\boldsymbol{\beta}
-{\boldsymbol{\mu}}_{0}) + 
(\boldsymbol{\beta} -\hat{\boldsymbol{\beta}}_{0})' \delta
\mathbf{X}_0'\mathbf{X}_{0} (\boldsymbol{\beta} -\hat{\boldsymbol{\beta}}_{0})=
Q(\delta, \boldsymbol{\beta}) + \delta b H_{0}(\delta),
$$ while to finish the third line we use the identity in (\ref{eq:identity2}).
Multiplying $\pi_{0}(\delta)\pi_{0}(\boldsymbol{\beta},\sigma^2)L(\boldsymbol{\beta},\sigma^2|D_0)^{\delta}$ by $C(\delta)^{-1}$ above yields the result (a). 
    
 \item[(b)] 
Since 
$$
Q(\delta, \boldsymbol{\beta}) + (\boldsymbol{\beta} -\hat{\boldsymbol{\beta}})' \mathbf{X}'\mathbf{X} (\boldsymbol{\beta} -\hat{\boldsymbol{\beta}}) = H(\delta) + {Q}^{*}(\delta,\boldsymbol{\beta}), 
$$
where $H(\delta)$ is defined in Theorem \ref{thm:LM} (b), and 
$${Q}^{*}(\delta,\boldsymbol{\beta}) = (\boldsymbol{\beta} - \boldsymbol{\mu}^{*})'(b\boldsymbol{R} + \delta \mathbf{X}_0'\mathbf{X}_{0} + \mathbf{X}'\mathbf{X})(\boldsymbol{\beta} - \boldsymbol{\mu}^{*}), $$
where $\boldsymbol{\mu}^{*} = (b\boldsymbol{R} + \delta \mathbf{X}_0'\mathbf{X}_{0} + \mathbf{X}'\mathbf{X})^{-1}[(b\boldsymbol{R} + \delta \mathbf{X}_0'\mathbf{X}_{0})\boldsymbol{\beta}^{*} + \mathbf{X}'\mathbf{X} \hat{\boldsymbol{\beta}}]$,  
using the normalized power prior in (a), the posterior $\pi(\boldsymbol{\beta}, \sigma^2, \delta | D_0, D)$ is of the form
$$
{\small 
\pi(\boldsymbol{\beta}, \sigma^2, \delta | D_0, D) \propto \frac{\pi_{0}(\delta) M_{0}(\delta)}{(\sigma^2)^{\frac{n+\delta n_0 + bk}{2}+a}}
\exp\left\{
-\frac{\delta\left[{S}_0 + b{H}_0(\delta)\right] +  S+ H(\delta) +
{Q}^{*}(\delta,\boldsymbol{\beta})}{2\sigma^2}\right\}. }
$$
Marginalizing $(\boldsymbol{\beta}, \sigma^2)$ out, we obtain   
\begin{align*}
\pi(\delta |D_0, D) &\propto \pi_{0}(\delta) M_{0}(\delta) \Gamma( \nu^{*}) 
|b\boldsymbol{R} +\delta \mathbf{X}_0'\mathbf{X}_0+ \mathbf{X}'\mathbf{X}|^{-\frac{1}{2}} \\ &\times\left\{\frac{\delta\left[{S}_0 + b{H}_0(\delta)\right] +  S+ H(\delta)}{2}\right\}^{-\nu^{*}},
\end{align*}
where $\nu^{*} = \frac{n+\delta n_0+(b-1)k}{2}+a-1$. Plugging in $M_{0}(\delta)$ we get (b). 

 \item[(c)] 
Integrating $\sigma^2$ out from the joint posterior, we have 
$$
{\footnotesize \pi(\boldsymbol{\beta}, \delta | D_0, D) \propto \pi_{0}(\delta)M_{0}(\delta)\Gamma\left(\nu^{*}+\frac{k}{2}\right)
\left\{\frac{\delta\left[{S}_0 + b{H}_0(\delta)\right] +  S+ H(\delta) + {Q}^{*}(\delta,\boldsymbol{\beta})}{2}\right\}^{-\left(\nu^{*}+\frac{k}{2}\right)},}
$$
where $\nu^{*}$ and ${Q}^{*}(\delta,\boldsymbol{\beta})$ are defined above in the proof of part (b). The conditional distribution of $\boldsymbol{\beta}$ given $(\delta, D_0, D)$  satisfies
\begin{align*}
\pi(\boldsymbol{\beta} | \delta, D_0, D) & \propto 
\left\{\delta\left[{S}_0 + b{H}_0(\delta)\right] +  S+ H(\delta) + {Q}^{*}(\delta,\boldsymbol{\beta})\right\}^{-\left(\nu^{*}+\frac{k}{2}\right)} \\
& \propto 
\left\{ 1 + \frac{1}{\nu} \left[ \frac{(\boldsymbol{\beta} - \boldsymbol{\mu}^{*})' \nu(b\boldsymbol{R} + \delta \mathbf{X}_0'\mathbf{X}_{0} + \mathbf{X}'\mathbf{X})(\boldsymbol{\beta} - \boldsymbol{\mu}^{*})}
{\delta\left\{{S}_0 + b{H}_0(\delta)\right\} +  S+ H(\delta) } \right] \right\}^{-\frac{\nu + k}{2}}, 
\end{align*}
where $\nu = (b-1)k+\delta n_0+n+2a-2.$
This is the kernel of a multivariate Student {\em t}-distribution  with parameters specified in Theorem \ref{thm:LM} (c).

\item[(d)] 
Using Gaussian integral we can marginalize $\boldsymbol{\beta}$ out from the joint posterior, then 
$${\small 
\pi(\sigma^2, \delta | D_0, D) \propto \frac{\pi_{0}(\delta) M_{0}(\delta)}{(\sigma^2)^{\nu^{*}+1}} 
\exp\left\{-
\frac{\delta\left[S_{0} + b H_{0}(\delta)\right] + S + H(\delta) }{2 \sigma^2} 
\right\} |b\boldsymbol{R} + \delta \mathbf{X}_0'\mathbf{X}_{0} + \mathbf{X}'\mathbf{X}|^{-\frac{1}{2}},}
$$
where $\nu^{*}$ is defined in the proof of part (b). Conditional on $(\delta, D_0, D)$, $\pi(\sigma^2 |\delta, D_0, D)$ is an inverse-gamma kernel with parameters specified in Theorem \ref{thm:LM} (d).
\end{itemize}

\noindent \textbf{Proof of Theorem \ref{th:optimality}:}

The quantity $L_g$ in (\ref{def:WeightedKL}) can be written as \begin{align}\label{eq:Lg}
L_g &= E_{\pi_{0}(\delta)}\left\{ (1-\delta) K(g,\pi_0) + \delta K(g,\pi_1)\right\}\nonumber \\\nonumber &= E_{\pi_{0}(\delta)}\left[ \int_{\boldsymbol{\Theta}} g(\boldsymbol{\theta}|\delta) \log\left\{\frac{g(\boldsymbol{\theta}|\delta)^{1-\delta}}{\pi_0(\boldsymbol{\theta})^{1-\delta}}\cdot \frac{g(\boldsymbol{\theta}|\delta)^{\delta}}{\pi_1(\boldsymbol{\theta})^{\delta}}\right\}d\boldsymbol{\theta}\right] \\ \nonumber&= E_{\pi_{0}(\delta)}\left[ \int_{\boldsymbol{\Theta}} g(\boldsymbol{\theta}|\delta) \log\left\{\frac{g(\boldsymbol{\theta}|\delta)}{Q(D_0)^{\delta}\pi_{0}(\boldsymbol{\theta})L(\boldsymbol{\theta}|D_0)^{\delta}}\right\}d\boldsymbol{\theta}\right] \\&= E_{\pi_{0}(\delta)}\left\{K[g(\boldsymbol{\theta}|\delta),
\pi^*(\boldsymbol{\theta}|\delta,D_0)]\right\} - E_{\pi_{0}(\delta)}\left[\log\left\{\frac{Q^{\delta}(D_0)}{Q_1(D_0,\delta)}\right\} \right],
\end{align}
where \begin{eqnarray}\label{eq:A2}
\pi^*(\boldsymbol{\theta}|\delta,D_0) = \frac{L(\boldsymbol{\theta}|D_{0})^\delta 
\pi_{0}(\boldsymbol{\theta}) }{\int_{\boldsymbol{\Theta}} 
L(\boldsymbol{\theta}|D_{0})^\delta \pi_{0}(\boldsymbol{\theta}) 
d\boldsymbol{\theta}},
\end{eqnarray} $Q(D_0)$ is defined in (\ref{eq:QD0}), and $Q_1(D_0,\delta)^{-1}$ is the denominator in (\ref{eq:A2}). 
The second term of (\ref{eq:Lg}) in the last line 
is not related to $g$, and the inside KL divergence in the first term is clearly minimized when  $g(\boldsymbol{\theta}|\delta)=\pi^*(\boldsymbol{\theta}|\delta,D_0)$.

\smallskip
\noindent \textbf{Proof of Theorem \ref{th:mode}:}

\noindent Applying the property of the KL  
divergence between two distributions,
\begin{equation*}
K(f_{1},f_{2})=\int f_{1}(x)\log \frac{f_{1}(x)}{f_{2}(x)}dx\geq 0,
\end{equation*}
with equality held if and only if $f_1(x)=f_2(x)$, we conclude that
\begin{align}
  \label{eq:KLIneq}
   &\frac{n}{n_0}h_1(D_0,D,\delta)=\int_{\boldsymbol{\Theta}} \log L(\boldsymbol{\theta}|D)\{
  \pi(\boldsymbol{\theta}|D_{0},D,\delta)-\pi(\boldsymbol{\theta}|D_{0},\delta)\}
  d\boldsymbol{\theta}
   \nonumber \\
   &=\int_{\boldsymbol{\Theta}} \log\left\{\frac{\pi(\boldsymbol{\theta}|D_0,D,
  \delta)}{\pi(\boldsymbol{\theta}|D_0,\delta)}M(D_0,D|\delta)\right\}\{
  \pi(\boldsymbol{\theta}|D_{0},D,\delta)-\pi(\boldsymbol{\theta}|D_{0},\delta)\}
  d\boldsymbol{\theta}
 \nonumber\\
  &=\int_{\boldsymbol{\Theta}} \log
\frac{\pi(\boldsymbol{\theta}|D_{0},D,\delta)}
{\pi(\boldsymbol{\theta}|D_{0},\delta)}\pi(\boldsymbol{\theta}|D_{0},D,\delta)
d\boldsymbol{\theta} + \int_{\boldsymbol{\Theta}} \log
\frac{\pi(\boldsymbol{\theta}|D_{0},\delta)}
{\pi(\boldsymbol{\theta}|D_{0},D,\delta)}\pi(\boldsymbol{\theta}|D_{0},\delta)
d\boldsymbol{\theta} \geq 0,
\end{align}
with equality held if and only if
$\pi(\boldsymbol{\theta}|D_{0},D,\delta)=\pi(\boldsymbol{\theta}|D_{0},\delta)$. 
In (\ref{eq:KLIneq}), $M(D_0,D|\delta)$ is a marginal density that does
not depend on $\boldsymbol{\theta}$ and hence its related term is 0 since both
$\pi(\boldsymbol{\theta}|D_{0},D,\delta)$ and $\pi(\boldsymbol{\theta}|D_{0},\delta)$ are
proper.

In order to show that the marginal posterior mode of $\delta$ is 1,
it is sufficient to show that the  derivative of $\pi(\delta|D_0,D)$
in (\ref{aposterior}) is non-negative. 
Using certain algebra similar to the proof of  
identity (\ref{eq:calpha}), we
obtain
\begin{align}\label{eq:DeltaDerivative}
  \frac{d}{d\delta}\pi(\delta|D_0,D)
  &=\frac{d}{d\delta}\{\log\pi_{0}(\delta)\} \pi(\delta|D_0,D)~ + \nonumber\\
  & \pi(\delta|D_0,D)\int_{\boldsymbol{\Theta}}\log L(\boldsymbol{\theta}|D_0) \{
  \pi(\boldsymbol{\theta}|D_{0},D,\delta)-
  \pi(\boldsymbol{\theta}|D_{0},\delta)\}d\boldsymbol{\theta}.
\end{align}

Since we are dealing with the exponential family with the form
(\ref{eq:expfamdensity}) and (\ref{eq:currentdensity}), 
considering the likelihood ratio we have
\begin{align}
  \label{eq:LikeliRatio}
  &\log L(\boldsymbol{\theta}|D_0) =\log h(D_0)
  +n_0\{\underline{T}(D_0)'\underline{w}(\boldsymbol{\theta})+\tau(\boldsymbol{\theta})\}
  \nonumber \\
  &=\log h(D_0)-\frac{n_0}{n}\log h(D) +\frac{n_0}{n}\log
  L(\boldsymbol{\theta}|D)
  +n_0\{\underline{T}(D_0)-\underline{T}(D)\}'\underline{w}(\boldsymbol{\theta}).
\end{align}

Combining (\ref{eq:KLIneq}) and (\ref{eq:LikeliRatio}) into
(\ref{eq:DeltaDerivative}), we prove Theorem \ref{th:mode} by
showing the condition (\ref{eq:Mode1Cond}).

\medskip

\noindent \textbf{Proof of Theorem \ref{th:Mode0Result}:}

\noindent Suppose that $k$ is an arbitrary positive constant. We
take the likelihood function of the form $L(\boldsymbol{\theta}|x)=k
f(x|\boldsymbol{\theta})$, then 
$L(\boldsymbol{\theta}|D)=k^{n} f(D|\boldsymbol{\theta})$ and 
$L(\boldsymbol{\theta}|D_{0})=k^{n_{0}} f(D_{0}|\boldsymbol{\theta})$. 
For the original joint power
prior,  the marginal posterior distribution of $\delta$ can
be rewritten as
\begin{align}\label{eq:OrigPriorMargDelta}
\pi(\delta|D_{0},D)& \propto \pi_{0}(\delta) \int_{\boldsymbol{\Theta}}
L(\boldsymbol{\theta}|D)L(\boldsymbol{\theta}|D_{0})^\delta \pi_{0}(\boldsymbol{\theta}) 
 d\boldsymbol{\theta}  \nonumber
\\&\propto \pi_{0}(\delta) \int_{\boldsymbol{\Theta}}
f(D|\boldsymbol{\theta})[k^{n_{0}}f(D_{0}|\boldsymbol{\theta})]^\delta 
\pi_{0}(\boldsymbol{\theta}) d\boldsymbol{\theta}.
\end{align}
To prove that the marginal posterior mode of $\delta$ is $0$, it
is sufficient to show that the derivative of 
$\pi(\delta|D_{0},D)$ with respect to $\delta$ 
is non-positive for any $\delta \in [0,1]$.

The derivative contains two parts.  The 
first part is the derivative on $\pi_{0}(\delta)$.  If $\pi_{0}(\delta)$ is
non-increasing as described in the theorem, this part is
non-positive.  The second part is the derivative in the integral
part in (\ref{eq:OrigPriorMargDelta}). 
An equivalent condition to guarantee this part non-positive is 
\begin{align}\label{appaProof}
&\int_{\boldsymbol{\Theta}}
f(D|\boldsymbol{\theta})\frac{d[k^{n_{0}}f(D_{0}|\boldsymbol{\theta})]^\delta}
{d\delta}
\pi_{0}(\boldsymbol{\theta})  d\boldsymbol{\theta} \leq 0 \nonumber \\
&\Longleftrightarrow k^{n_{0}\delta} \int_{\boldsymbol{\Theta}}
\pi_{0}(\boldsymbol{\theta})f(D|\boldsymbol{\theta})f(D_{0}|\boldsymbol{\theta})^\delta 
\{n_{0}\log k+\log f(D_{0}|\boldsymbol{\theta})\} d\boldsymbol{\theta} \leq 0 \nonumber \\
&\Longleftrightarrow \frac{\int_{\boldsymbol{\Theta}}
\pi_{0}(\boldsymbol{\theta})f(D|\boldsymbol{\theta})f(D_{0}|\boldsymbol{\theta})^\delta \log
f(D_{0}|\boldsymbol{\theta})d\boldsymbol{\theta}}{\int_{\boldsymbol{\Theta}}
\pi_{0}(\boldsymbol{\theta})f(D|\boldsymbol{\theta})
f(D_{0}|\boldsymbol{\theta})^\delta d\boldsymbol{\theta}} \leq
n_{0}\log\frac{1}{k},
\end{align}
assuming that the derivative and integral are interchangeable.

If we take
\begin{equation*} k_{0}=\exp\bigg\{
-\frac{1}{n_0}\underset{0\le\delta\le 1}{\max}\frac{\int_{\boldsymbol{\Theta}}
\pi_{0}(\boldsymbol{\theta})f(D|\boldsymbol{\theta})f(D_{0}|\boldsymbol{\theta})^\delta
\log f(D_{0}|\boldsymbol{\theta})d\boldsymbol{\theta}}{\int_{\boldsymbol{\Theta}}
\pi_{0}(\boldsymbol{\theta})f(D|\boldsymbol{\theta})
f(D_{0}|\boldsymbol{\theta})^\delta d\boldsymbol{\theta}} \bigg\}>0,
\end{equation*}
then the sufficient condition in (\ref{appaProof}) for the
marginal posterior mode of $\delta$ being $0$ is met for any
$\delta$.

\section{MCMC Sampling Scheme}\label{appb}

\subsection{Algorithm for Posterior Sampling}
\label{app-algorithm1}

Here we describe an algorithm in detail that is 
applicable in models when  
$\pi(\delta|\boldsymbol{\theta}, D_0, D)$ is 
free of any numerical integration, and 
the full conditional for each $\theta_i$ is 
readily available. 

Let $\boldsymbol{\theta} = (\theta_1, \ldots, \theta_k)$ 
denote the parameters of interest in the model, and $\boldsymbol{\theta}_{-i}$ is 
$\boldsymbol{\theta}$ with the $i^{th}$ element removed. 
The initial prior $\pi_{0}(\boldsymbol{\theta})$ 
can be chosen so that the full 
conditional posterior of each $\theta_i$, the 
$\pi(\theta_i|\boldsymbol{\theta}_{-i}, \delta, D_0, D)$,
can be sampled 
directly using the Gibbs sampler \citep{Gelman13}. 
However, neither the full conditional posterior 
$\pi(\delta|\boldsymbol{\theta}, D_0, D)$ 
nor the marginal posterior $\pi(\delta|D_0, D)$ 
is readily available. Given that $\pi(\delta|D_0, D)$ is known up to a 
normalizing constant, the Metropolis-Hastings algorithm \citep{Chib95} is implemented. 
Here we illustrate the use of a random-walk Metropolis-Hastings algorithm 
with Gaussian proposals for $\vartheta = {\rm logit}(\delta)$, 
which converges well empirically. Let 
$q(\cdot ~|~\delta^{\rm old})$ denotes the proposal distribution for $\delta$ 
in the current iteration, given its value in the previous iteration is $\delta^{\rm old}$.
The algorithm proceeds as follows:

\begin{itemize}

\item[Step 0:] 
Choose the initial values for the parameters ${\boldsymbol{\theta}}^{(0)}$ 
and ${\delta}^{(0)}$, 
set the tuning constant as $c$, and iteration index $l=0$.

\item[Step 1:] The Metropolis-Hastings step. 
Simulate ${\vartheta}^* \sim {\rm N}({\vartheta}^{(l)}, c)$ and $U \sim {\rm unif}(0,1)$. 
Compute ${\delta}^{*} = {\rm logit}^{-1}(\vartheta^{*})$ and 
the acceptance probability $\alpha = \min\{1,t\}$. 
After applying a change of variable, we have
\begin{equation}
t  = 
\frac{\pi(\delta^{*} ~|~ D_0, D) 
q({\delta}^{(l)}~|~{\delta}^*) }
{\pi(\delta^{(l)} ~|~ D_0, D)  
q({\delta}^{*}~|~{\delta}^{(l)}) } = 
\frac{\pi(\delta^{*} ~|~ D_0, D) 
\delta^{*}(1-\delta^{*}) }
{\pi(\delta^{(l)} ~|~ D_0, D)  
\delta^{(l)}(1-\delta^{(l)}) }.
\nonumber \\
\end{equation}
 
Then set ${\delta}^{(l+1)} = \delta^*$, if $U < \alpha$. 
Otherwise, set ${\delta}^{(l+1)} = {\delta}^{(l)}$. 

\item[Step 2:] The Gibbs sampling step. For $i = 1,\ldots, k$, independently
sample  $\theta_i^{(l+1)}$ from its full conditional posterior 
$\pi(\theta_i|\boldsymbol{\theta}_{-i}^{(l)}, \delta^{(l+1)}, D_0, D)$. 

\item[Step 3:] 
Increase $l$ by $1$, and repeat steps $1$ and $2$ until the states have reached the 
equilibrium distribution of the Markov chain.

\end{itemize}

Since $\delta \in [0,1]$, 
an independent proposal from a beta distribution might also 
provide good convergence. 
In such cases, the proposal distribution $q(\cdot)$ will 
be the {\em same} beta distribution evaluated at $\delta^{(l)}$ and
$\delta^{*}$ in the nominator and denominator respectively.

\subsection{Algorithm to Compute the 
Scale Factor} \label{app-algorithm2}

Here we describe an algorithm in detail when the scale factor 
in the denominator, 
$C(\delta) = \int_{\boldsymbol{\Theta}} L(\boldsymbol{\theta}|D_{0})^\delta 
\pi_{0}(\boldsymbol{\theta}) 
d\boldsymbol{\theta}$ 
needs to be calculated numerically. 
From identity (\ref{eq:calpha}), 
$ 
\log C(\delta) = \int_{0}^{\delta} 
E_{\pi(\boldsymbol{\theta} | D_0, \delta^* )} 
\{ \log [ L( \boldsymbol{\theta}|D_0 ) ] \}  d {\delta^*}$, 
so we only need to calculate the one-dimensional integral.

MCMC samples from $\pi(\boldsymbol{\theta}|D_{0},\delta)$ 
with fixed $\delta$
can be easily drawn, since the target density 
is expressed explicitly up to a normalizing constant. 
A fast implementation with {\tt RStan} \citep{Carpenter17} and 
parallel programming is applicable, 
by including the fixed $\delta$ in the target statement. 
We develop the following 
algorithm to calculate the scale factor 
$\log C(\delta)$ up to a true constant. 
It is an adaptive version of the path sampling 
based on the results in \cite{Rosmalen18}.

\begin{itemize}

\item[Step 0:] 
Choose a set of $n-1$ different numbers as knots 
between $0$ and $1$, and another knot at $1$, with $n$ sufficiently large. 
Sort them in ascending order $(\delta_1, \ldots, \delta_{n-1}, 1)$. 
Let $\Delta_1 = \delta_1$, $\Delta_i = \delta_i - \delta_{i-1}$ 
($1 <i \leq n$), and $\Delta_{n} = 1-\delta_{n-1}$. 
Choose $M$, the number of MCMC samples 
in a run when sampling from 
$\pi(\boldsymbol{\theta}|D_{0},\delta)$.
Initialize $l = 1$. 

\item[Step 1:] 
Generate $M$ samples from 
$\pi(\boldsymbol{\theta}|D_{0},\delta_{l})$ using an appropriate 
MCMC algorithm. 
Denote the sample as $(\boldsymbol{\theta}^{(1)}_{l}, \boldsymbol{\theta}^{(2)}_{l}, \ldots, 
\boldsymbol{\theta}^{(M)}_{l} ) $.

\item[Step 2:]
Calculate 
$h(\delta_{l}) = \sum_{j=1}^{M} 
\log L( \boldsymbol{\theta}^{(j)}_{l}|D_0 )/M$.

\item[Step 3:]
Calculate 
$\log C(\delta_{l}) \approx 
\sum_{k=1}^{l} \Delta_{k} h(\delta_{k}) $.

\item[Step 4:]
Increase $l$ by $1$. If $l \leq n$ then 
repeat Steps 1 to 3. 
\end{itemize}

The output is 
a vector of $n$ values,  
$( \log C(\delta_{1}), \ldots, \log C(\delta_{n-1}), \log C(1))$, 
for selected knots. 

Finally, for $\delta$ that is not on the knots, 
it is efficient to 
linearly interpolate $\log C(\delta)$ based on its nearest 
two values 
on the knots \citep{Rosmalen18}. 
The interpolation can be done quite fast at every iteration when 
sampling from the posterior 
$\pi(\boldsymbol{\theta}, \delta | D_0, D)$ using a 
normalized power prior, so the algorithm similar to the one 
described in \ref{app-algorithm1} can be applied. 
Compared to the joint power prior, the 
extra computational cost is to calculate $\log C(\delta)$ 
on the selected knots, with the capability of parallel computation. 
Both of the algorithms in \ref{app-algorithm1} and 
\ref{app-algorithm2} are implemented in  {\tt R} package {\tt NPP}.

\end{document}